\documentclass[a4paper,11pt]{article}
\pdfoutput=1 

\usepackage{jinstpub} 

\usepackage{lineno}
\usepackage{subcaption}

\title{Geant4Reweight: a framework for evaluating and propagating hadronic interaction uncertainties in Geant4}

\author[a,1]{J. Calcutt\note{Corresponding author.}}
\author[b]{C. Thorpe}
\author[a]{K. Mahn}

\author[c]{L. Fields}

\affiliation[a]{Michigan State University, Department of Physics and Astronomy, East Lansing, Michigan, USA}
\affiliation[b]{Lancaster University, Department of Physics, LA1 4YB, Lancaster, U.K.}
\affiliation[c]{University of Notre Dame, Department of Physics, Notre Dame, Indiana, USA}

\emailAdd{calcuttj@msu.edu}

\abstract{Geant4Reweight is an open-source C++ framework that allows users to weight tracks produced by the Geant4 particle transport Monte Carlo simulation according to hadron interaction cross section variations and estimate uncertainties in Geant4 interaction models by comparing the simulation's hadron interaction cross section predictions to data. The ability to weight hadron transport as simulated by Geant4 is crucial to the propagation of systematic uncertainties related to secondary hadronic interactions in current and upcoming neutrino oscillation experiments, including MicroBooNE, NOvA, and DUNE, as well as hadron test beam experiments such as ProtoDUNE. We provide motivation for weighting hadron tracks in Geant4 in the context of systematic uncertainty propagation, a description of Geant4's transport simulation technique, and a description of our weighting technique and fitting framework in the momentum range 0--10 GeV/c, which is typical for the hadrons produced by neutrino interactions in these experiments. }

 
\begin{document}
\maketitle
\flushbottom

\section{Introduction}
\label{sec:intro}

Measurements of neutrino oscillation parameters and neutrino-nucleus ($\nu$N) cross sections performed by neutrino experiments, such as T2K~\cite{t2k_experiment}, MicroBooNE~\cite{uboone}, NOvA~\cite{nova}, and the upcoming experiment DUNE~\cite{dune}, rely on determining the energy, flavor, sign,\footnote{Whether it is a neutrino or antineutrino} and interaction topology of the detected neutrino event. However, these experiments can only infer each of these through the identification of $\nu$N interaction products. Issues in identifying these observables can arise in various and subtle ways, many of which are associated with the re-interaction in the detector of hadrons produced by the $\nu$N interaction --- so called ``secondary interactions'' (SI). A few examples are given below.  


\begin{enumerate}
    \item Some experiments, such as NOvA and (in the future) DUNE, reconstruct the neutrino energy via calorimetry of the neutrino interaction products. For example, consider a $\nu_\mu$ Charged Current (CC) interaction of the form $\nu_\mu + N \rightarrow  \mu^- + p + \pi^+ + N'$. Here, the energy of the hadronic system can be reconstructed using all non-muon energy depositions in the detector and by identifying the $\pi^+$ track and including its rest mass. The $\pi^+$ can undergo SI wherein it is absorbed by a nearby nucleus in the detector i.e. $\pi^+ + N \rightarrow N' + 2n$. In this case,  a majority of the identifiable kinetic energy of the  $\pi^+$ as well as its rest mass (139.57 MeV) would be lost.
    
    \item Experiments such as T2K use estimators for neutrino energy which rely only on lepton kinematic information, and assume the event was a CCQE interaction ($\nu_\mu + p \rightarrow \mu^- + n)$ within a nucleus. For interactions wherein a pion is produced, this energy estimate is biased. T2K identifies as CC0$\pi$ interactions events where there is no pion in the final state, and any number of protons,  to select a CCQE-enhanced event sample. Backgrounds within a CC0$\pi$ selection include CC1$\pi$ events --- which could arise from a resonant interaction ($\nu_\mu + p \rightarrow \mu^- + \Delta^{++} \rightarrow \mu^- + p + \pi^+$) --- wherein the outgoing $\pi^+$ is absorbed nearby the neutrino interaction due to SI.
    
    \item The signal for a $\nu_e$ CC event ($\nu_e + N \rightarrow e^- + p + N'$) includes an electromagnetic cascade produced by the final state $e^-$ (also known as a ``shower"). The electron in the $\nu_e$ CC event is identified by the presence of one such shower. Backgrounds arise from the presence of a $\pi^0$ in the final state, as this particle promptly decays to photons that can produce similar showers. This $\pi^0$ could be produced as a product of SI in the following interaction $\nu_\mu + N \rightarrow \nu_\mu + p + \pi^+$ if the $\pi^+$ undergoes a charge exchange interaction ($\pi^+ + N \rightarrow \pi^0 + N'$) close to the $\nu$N vertex. 
    
    \item Consider a Neutral Current interaction of the form $\bar{\nu}_\mu + N \rightarrow \bar{\nu}_\mu + p + \pi^+ + N'$. In the case that the $\pi^+$ is particularly energetic and does not undergo SI, it could mimic a $\mu$ as it comes to stop in the detector. The presence of the apparent $\mu$ would potentially cause this event to be selected as a CC event, meaning this event would contribute to a background. 
\end{enumerate}

Experiments typically correct for these effects in data using estimates from Monte Carlo simulation. However, mismodeling of hadron SI rates will lead to biases in this correction. An underprediction of hadron SI would underestimate the neutrino energy (example 1) or overestimate the $\nu_e$-CC event rate (example 3). An overestimation of $\pi^+$ SI would underestimate the rate of backgrounds described in example 4. These examples highlight two needs from experiments: 1) the ability to estimate uncertainties related to hadron SI and 2) the ability to propagate this uncertainty to signal and background event rates.

Neutrino experiments typically rely on the Geant4~\cite{geant4_1}\cite{geant4_2}\cite{geant4_3} toolkit to simulate particles resulting from $\nu$N interactions as they pass through detectors. 
This toolkit achieves this with the use of particle-nuclei cross sections (to determine the rates of interactions occurring in the detector) and physics models (to determine the specific dynamics of these interactions).
Prior to running, users choose which cross sections and models are used in the simulation by selecting from standard or custom ``physics lists''. Additionally, certain model parameters can be varied before run-time. If users intend to produce varied simulation results by choosing from a series of physics lists or tweaking model parameters (i.e. for the purpose of uncertainty propagation), they must rerun the full simulation and downstream processing chain multiple times.
This is computationally inefficient, especially when considering any further simulation required for a neutrino experiment (i.e. detector response and reconstruction).
A common alternative to this is an event-by-event weighting scheme which requires only one ensemble of Monte Carlo events to be produced. This is commonly referred to as "reweighting." When reweighting, the MC simulation is generated once under some nominal model. The results of each simulated event are then considered under some varied model, and weights are given to the events. When applied, the weights will modify the population of events throughout regions of phase space simulated using the varied model, thus approximating the results of rerunning the simulation under the varied model. 
Independent implementations of hadronic interaction reweighting have  been used by recent neutrino experiments to propagate hadron interaction uncertainties to their neutrino flux predictions~\cite{T2K_Flux}\cite{MINERvA_Flux}. A similar weighting technique to that described in this paper has also been used by T2K to propagate hadron SI model uncertainties~\cite{T2K_NuE_Appearance}.
In the context of neutrino interactions, similar weighting schemes exist to propagate systematic uncertainties on particle rescattering within the nuclear medium (so-called Final State Interactions)~\cite{DobsonReweighting}\cite{DobsonThesis}. 


The software package described here, Geant4Reweight~\cite{g4rw_git}, provides a framework in which Geant4's hadron interaction cross sections and models can be compared to data in order to estimate uncertainties, which can then be applied to analyses through reweighting. This framework is designed to be used in the momentum range of 0--10 GeV/c which is typical of hadrons produced by neutrino interactions in the aforementioned experiments. It is an open-source C++ framework that is intended to provide a unified approach across experiments to estimating and propagating systematic uncertainties related to hadron SI. Many internal methods rely on the ROOT~\cite{root} data analysis framework for storage or fitting routines. The rest of this paper describes the following: section \ref{sec:geant4} focuses on the technique used by Geant4 to simulate particle transport through material. This is described in order to inform the reweighting technique given in section \ref{sec:reweighting}. Section \ref{sec:example} then provides examples of reweighting on simulated $\pi^+$ tracks. Section \ref{sec:fitting} describes the fitting machinery used to estimate uncertainties on Geant4 cross sections by comparing to external hadron scattering data.


\section{Geant4 simulation strategy}
\label{sec:geant4}
Before discussing the technical details of this reweighting framework, this section provides a discussion of Geant4's transport simulation. A detailed description of Geant4's simulation is available in Ref~\cite{geant4_app}. This section provides a short description of the portion relevant to hadron SI reweighting.   

\begin{figure}
    \centering
    \includegraphics[width=.5\textwidth]{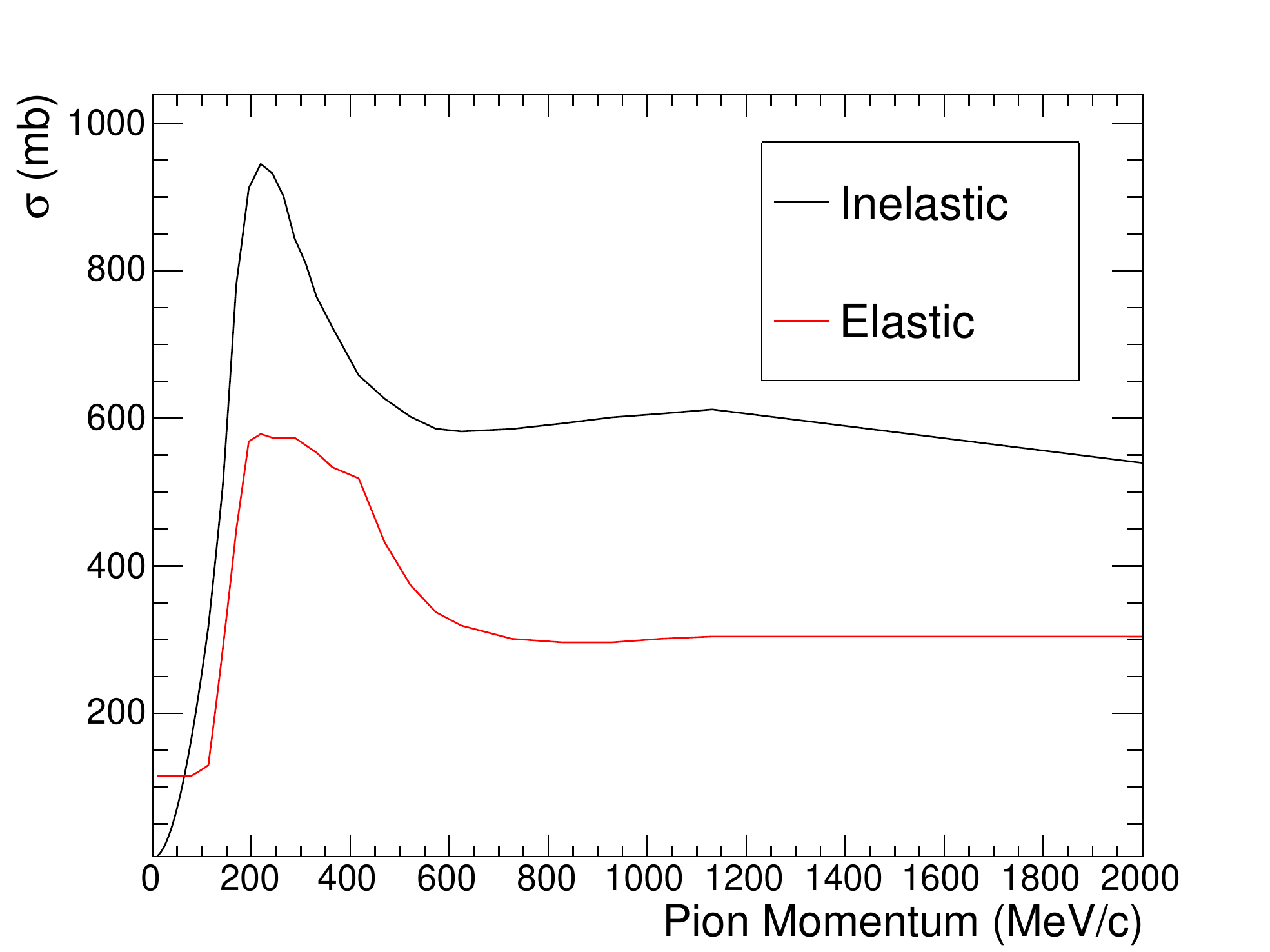}
    \caption{Geant4 cross section predictions for $\pi^+$-Ar interactions.}
    \label{fig:xsec}
\end{figure}

Within Geant4's simulation, a particle moves through the detector in an iterative process whereby it takes a set of steps between points in space and time. 
At each step, a set of processes (described further in the next paragraph)\footnote{``Process'' is the technical term for what is happening to the particle during that step --- i.e. it is simply moving between points or it is undergoing an interaction.} are considered and the simulation decides which processes are invoked. The processes are categorized into three types (\textit{at-rest}, \textit{continuous}, and \textit{discrete}) that determine when in the step they occur, and what criteria must be met before the process is invoked.\footnote{Some processes are hybrid processes with combined behavior of multiple types of processes. For example, decay processes are both \textit{discrete} and \textit{at-rest}. For our purposes, the two behaviors can be treated separately} If a particle stops, the \textit{at-rest} processes each propose an interaction time as an analog for an interaction length, and the process that proposes the shortest time is invoked. When a particle is not at rest, each \textit{continuous} and \textit{discrete} process proposes an interaction length, and the shortest one is chosen as the step length.\footnote{If the distance to the nearest volume boundary is shorter than the proposed interaction length, it is chosen as the step size.} Then, as the particle moves between step end-points, each \textit{continuous} process is invoked. Finally, at the end of the step, if a \textit{discrete} process proposed the shortest interaction length, that process is invoked. For processes we are considering in the context of the reweighting framework, the interaction lengths are provided by cross section tables in the case of inelastic and elastic hadron scattering and Coulomb scattering, and by lifetime tables in the case of pion decays. The inelastic and elastic cross sections for $\pi^+$ interacting on argon are shown in figure \ref{fig:xsec}. These cross sections at this momentum range are taken from calculations performed by Barashenkov~\cite{geant4_phys}~\cite{barashenkov}.

When a process is chosen to occur, Geant4 invokes an implementation of the model corresponding to that process. The choice of model is customizable and set based on the physics list chosen by the user. The results of the model are a set of Geant4 tracks including the updated track (either with a change in momentum direction and energy or it is removed from the stack of active particles) after undergoing the process and any secondary tracks created by the process. Geant4Reweight is intended to treat hadrons in the range of 0--10 GeV/c 
using physics lists commonly used by neutrino experiments. The physics lists ``G4HadronPhysicsQGSP\_BERT'' and ``G4HadronElasticPhysics'' are commonly used by neutrino oscillation experiments at this momentum range for the inelastic and elastic hadronic interactions, and are what we show in this paper. For inelastic interactions, the Bertini Intranuclear Cascade model is used for hadrons up to 10 GeV/c~\cite{geant4_phys}.  The differential elastic cross section is calculated using the Glauber model~\cite{geant4_phys}\cite{glauber} above 1 GeV/c and the Gheisha model~\cite{gheisha} below that.

\section{Reweighting technique}
\label{sec:reweighting}
The previous section provided a description of the Geant4 simulation strategy and introduced relevant concepts such as processes, cross sections, interaction lengths, and models in the parlance of Geant4's simulation. This section describes the reweighting technique used to create weights for particle tracks under some given change to a process' cross section. The processes we are interested in reweighting (the hadronic elastic and inelastic interactions) are \textit{discrete} processes, so we will focus the remaining discussion on this type of process. To inform the description of the reweighting process, it is helpful to construct event probabilities that can be varied according to the cross sections. The following discussion describes this. 

When the particle track is first created, each \textit{discrete} process samples the exponential distribution:
\begin{equation}
p(N) = e^{-N}dN.
\end{equation}
This provides the process with a number of interaction lengths ($N$). The interaction length proposed for a step is $N\lambda$, where $\lambda$ is the mean free path of the process for that step and is inversely proportional to the cross section for that process. At each step, if the process in question did not occur, the process subtracts an amount equal to $n = \Delta L/\lambda$ from $N$ where $\Delta L$ is the step length taken and $n$ is the number of interaction lengths taken by the step. Once this process is invoked in the procedure detailed in section \ref{sec:geant4}, $N$ is sampled again. Thus, for a given process, the probability for that process to \textit{not} occur after a series of steps ($P_S$ in short for $P_{Survive}$) is given by the probability to sample $N$ greater than the sum of all interaction lengths traveled by particle ($N_T$):
\begin{equation}\
\begin{split}
   P_{S} &= P(N > N_T) 
   \\
   &= \int_{N_{T}}^{\infty}e^{-N}dN = e^{-N_{T}} = e^{-\sum\limits_i^{f} n_{i}} 
   \\
   &= e^{-\sum\limits_i^{f} \Delta L_i/\lambda_i} = e^{-\sum\limits_i^{f} \Delta L_i\sigma_i}.
\end{split}
\end{equation}
where $n_i = \Delta L_i/\lambda_i$ is the number of interaction lengths traveled in step $i$. The last term uses the cross section $\sigma_i = 1/\lambda_i$ to simplify notation.
Normally, $1/\lambda = \sigma\rho N_A/M$, where $M$ and $\rho$ are the molar mass and density of the surounding material, and $N_A$ is Avogadro's number. The extra factors have been absorbed into $\sigma$ for clarity.
Note that the cross section $\sigma_i$ is generally dependent on the particle's momentum and the surrounding material for step $i$. This probability can be rewritten as the product of the probabilities for each step to occur without interaction:
\begin{equation}\label{eqn:p_survive}
    P_{S} = \prod\limits_i^{f}P_{S, i} = \prod\limits_i^{f}e^{-\Delta L_i\sigma_i}.
\end{equation}
If the process does occur, it only occurred in the last step, so the probability ($P_I$ in short for $P_{Interact}$) is largely unchanged except for the term contributing to the final step:
\begin{equation}\label{eqn:p_interact}
    \begin{split}
    P_{I} &= \bigg(\prod\limits_i^{f - 1}P_{S,i}\bigg)\bigg(P_{I, f}\bigg) = \bigg(\prod\limits_i^{f - 1}P_{S,i}\bigg)\bigg(1 - P_{S, f}\bigg) 
    \\
    &= \bigg(\prod\limits_i^{f - 1}e^{-\Delta L_i\sigma_i}\bigg)\bigg(1 - e^{-\Delta L_f\sigma_f}\bigg).
    \end{split}
\end{equation}

These probabilities can be easily extended to consider multiple processes. In equations \ref{eqn:p_survive} and \ref{eqn:p_interact}, the cross sections will be extended to be sums of the cross sections of all processes. 
\begin{equation}\label{eq:total_sigma}
    \sigma = \sigma_I + \sigma_E + \sigma_X
\end{equation}
For our purposes, we are interested in reweighting the elastic and inelastic hadronic interactions, so the cross sections for these processes are explicitly stated in equation \ref{eq:total_sigma} as $\sigma_I$ and $\sigma_E$ respectively. All other processes are combined into a single effective cross section $\sigma_X$. The processes included in $\sigma_X$ are Coulomb scattering and decay (for the relevant particles) which is effectively treated as an interaction with an interaction length given by $1/(\gamma \beta c \tau)$. Thus, for a given process $p$, the probability it will occur is given by
\begin{equation}\label{eqn:p_interact_spec}
    P_{I, p} = \bigg(\prod\limits_i^{N_f - 1}e^{-\Delta L_i\sigma_i}\bigg)\bigg(1 - e^{-\Delta L_f\sigma_f}\bigg)\bigg(\frac{\sigma_{p, f}}{\sigma_f}\bigg).
\end{equation}



\subsection{Weight calculations}
The weights are calculated by considering a change to the underlying cross sections and then determining whether the event was \textit{more} or \textit{less} probable under that change. Geant4Reweight has the ability to reweight both the inelastic and elastic hadronic cross sections ($\sigma_I$ and $\sigma_E$ respectively). 
We consider general variations to the inelastic and elastic cross sections as such:
\begin{equation}
\begin{split}
    \sigma_I &\rightarrow \sigma_I' = f_I\sigma_I
    \\
    \sigma_E &\rightarrow \sigma_E' = f_E\sigma_E.
\end{split}
\end{equation}
A weight is given to an entire track by considering each step, creating a weight for each step, and then multiplying these together:
\begin{equation}
W_{Track} = \prod\limits_i^{f}W_i.
\end{equation}
For each step, one of four things can happen:
\begin{enumerate}
    \item The particle undergoes no interaction
    \item The particle undergoes an inelastic interaction
    \item The particle undergoes an elastic interaction
    \item The particle undergoes some other interaction
\end{enumerate}
Each of these cases have different forms of weights. For the first, the weight is given by the change in the survival probability for the step:
\begin{equation}
    \begin{split}
    W_{S, i} &= e^{-\Delta L_i(\sigma_{I, i}' + \sigma_{E, i}' + \sigma_{X, i})}/e^{-\Delta L_i(\sigma_{I, i} + \sigma_{E, i} + \sigma_{X, i})}
    \\
    &= e^{-\Delta L_i(\sigma_{I, i}' + \sigma_{E, i}')}/e^{-\Delta L_i(\sigma_{I, i} + \sigma_{E, i})}
    \end{split}.
\end{equation}
The other three weights are given by changes to interacting probabilities for that step, as in the last two terms in equation \ref{eqn:p_interact_spec}.
\begin{equation}
    \begin{split}
    W_{I/E, i} &= \bigg(\frac{\sigma_{I/E, i}'}{\sigma_i'}\bigg)
                 \bigg(\frac{\sigma_i}{\sigma_{I/E, i}}\bigg)
                 \bigg(\frac{1 - e^{-\Delta L_i\sigma_i'}}{1 - e^{-\Delta L_i\sigma_i}}\bigg)
                 \\
              &= f_{I/E}\bigg(\frac{\sigma_i}{\sigma_i'}\bigg)
                 \bigg(\frac{1 - e^{-\Delta L_i\sigma_i'}}{1 - e^{-\Delta L_i\sigma_i}}\bigg)
    \end{split}
\end{equation}
\begin{equation}
    \begin{split}
    W_{X, i} &= \bigg(\frac{\sigma_{X, i}}{\sigma_i'}\bigg)
             \bigg(\frac{\sigma_i}{\sigma_{X, i}}\bigg)
             \bigg(\frac{1 - e^{-\Delta L_i\sigma_i'}}{1 - e^{-\Delta L_i\sigma_i}}\bigg)
             \\
          &= \bigg(\frac{\sigma_i}{\sigma_i'}\bigg)
              \bigg(\frac{1 - e^{-\Delta L_i\sigma_i'}}{1 - e^{-\Delta L_i\sigma_i}}\bigg)
    \end{split}                 
\end{equation}
Here, $\sigma_i = \sigma_{I, i} + \sigma_{E, i} + \sigma_{X, i}$ and $\sigma_i' = \sigma_{I, i}' + \sigma_{E, i}' + \sigma_{X, i}$.


\subsection{Exclusive channels}\label{subsec:exc_chan}
In neutrino experiments it is often important to consider not only the total (also called inclusive) inelastic hadronic interactions but also exclusive interactions. The examples listed in section \ref{sec:intro} highlight this. For examples 1 and 2, the instigating interaction is $\pi^+$ absorption. In example 3, it is $\pi^+$ charge exchange. However, in Geant4 the concept of exclusive hadronic interactions is not explicitly defined, and the only accessible information is the outgoing particles and their kinematics after a model is invoked. In this way, the model can be thought of as a ``black box'' that determines the results of the interaction (outgoing particles) which are then added to the simulation stack and whose information users are able to access. Therefore, Geant4Reweight defines exclusive interaction channels by the outgoing particles following an interaction. For pions, the exclusive channel names and their definitions are given in table \ref{tab:pi_exc}.

\begin{table}[htbp]
\centering
\caption{\label{tab:pi_exc} Summary of exclusive channel definitions for charged pions. $X$ represents any number of nucleons in the final state. $n\pi$ represents more than one $\pi$ (any charge) in the final state.}
\smallskip
\begin{tabular}{|l|l|}
\hline
Channel Name & Definition \\
\hline
Quasielastic & $\pi^{\pm} + N \rightarrow \pi^{\pm} + N' + X$\\
Absorption & $\pi^{\pm} + N \rightarrow N' + X$\\
Charge Exchange & $\pi^{\pm} + N \rightarrow \pi^0 + N' + X$\\
Double Charge Exchange & $\pi^{\pm} + N \rightarrow \pi^{\mp} + N' + X$\\
Production & $\pi^{\pm} + N \rightarrow n\pi + N' + X$\\
\hline
\end{tabular}
\end{table}

\begin{figure}[htbp]
    \centering
    \begin{subfigure}[t]{0.45\textwidth}
        \centering
        \includegraphics[width=\textwidth]{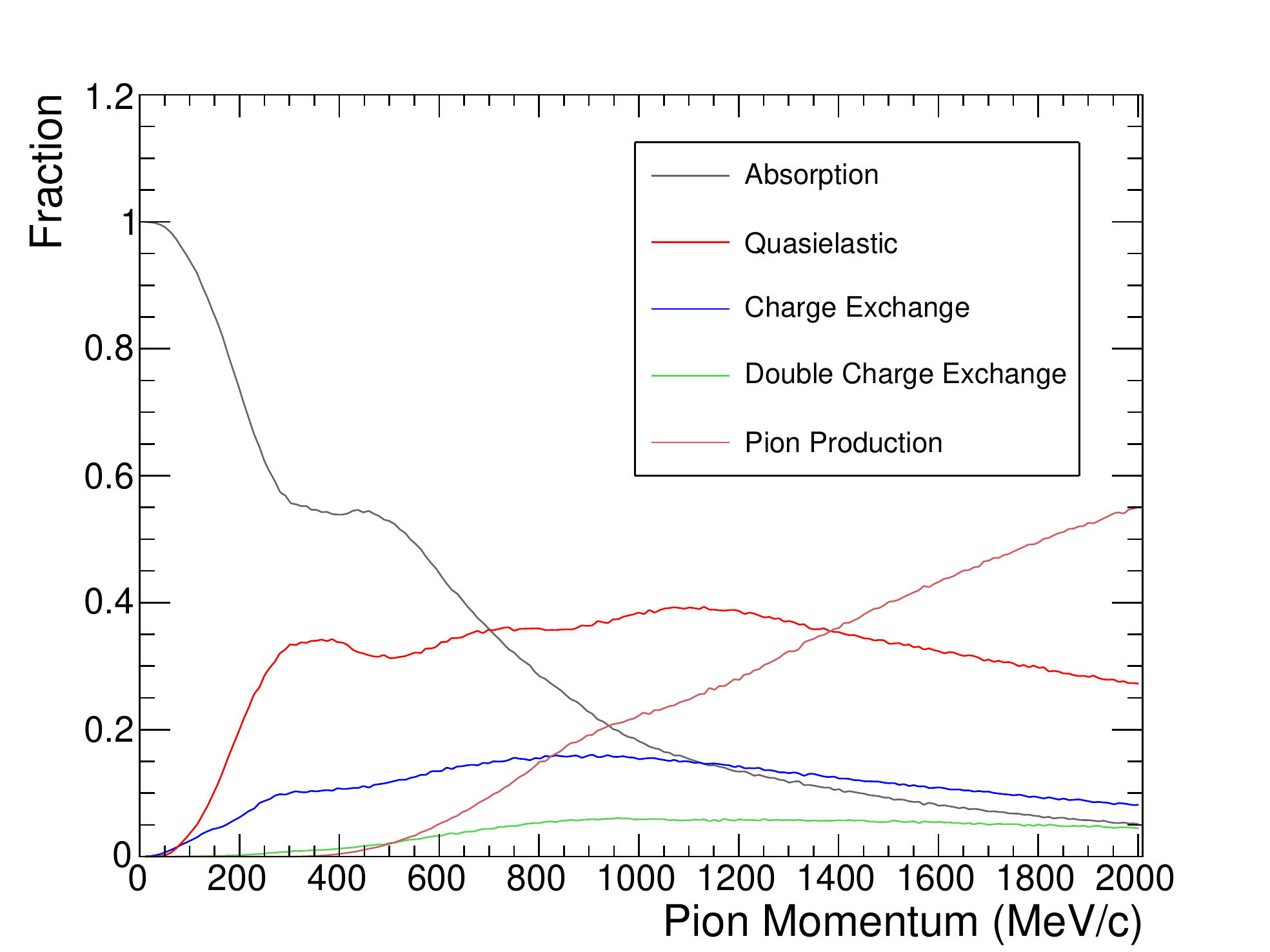}
        \caption{Exclusive final state fractions.}
    \end{subfigure}
    ~
    \begin{subfigure}[t]{0.45\textwidth}
        \centering
        \includegraphics[width=\textwidth]{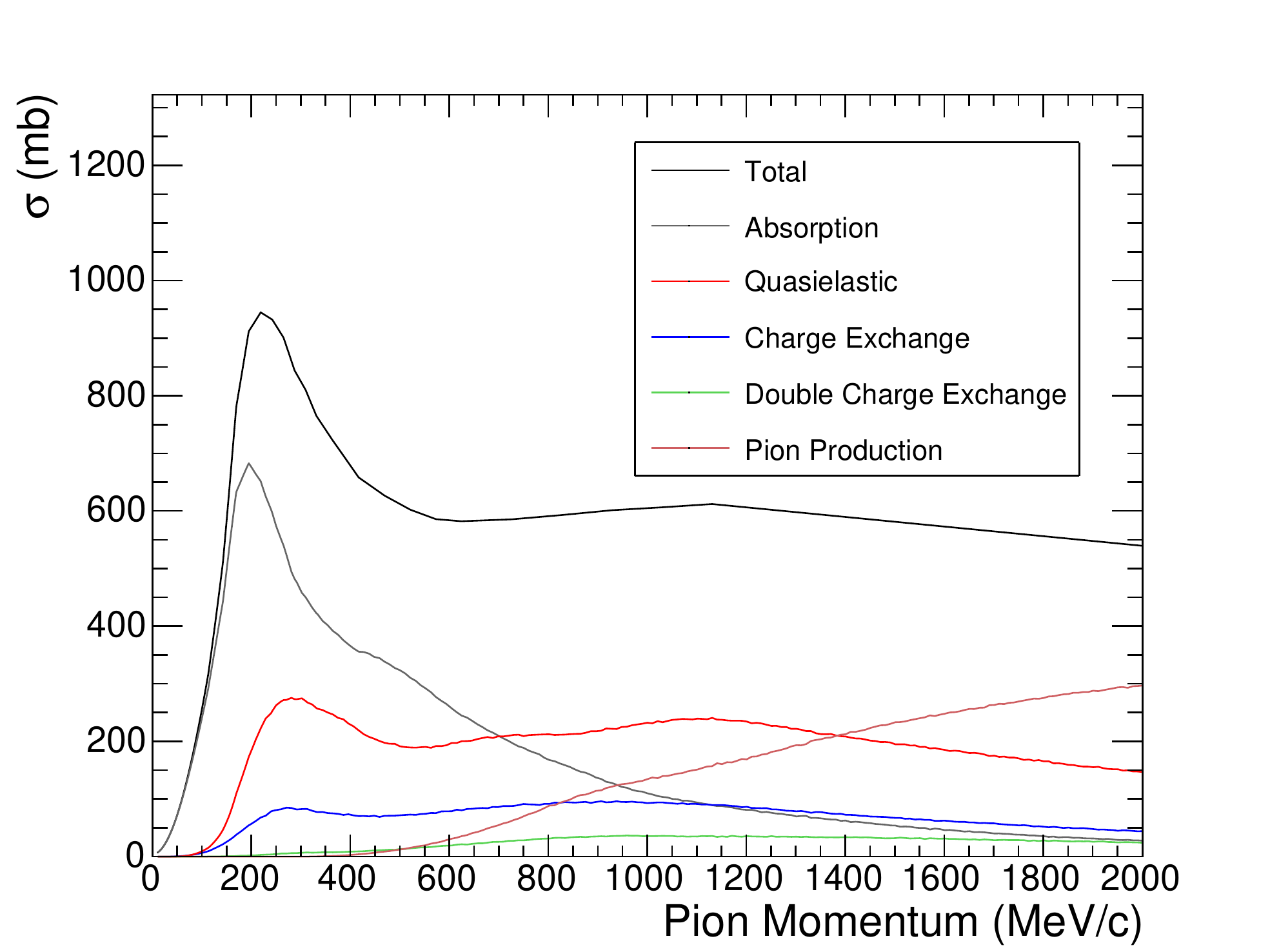}
        \caption{Exclusive \& Total Inelastic cross sections}
    \end{subfigure}
    \caption{Bertini Cascade model predictions and Geant4 cross section predictions for inelastic $\pi^+$-Ar interactions. The exclusive cross sections on the right are produced by multiplying the fractions on the left by the total inelastic cross section on the right.}
    \label{fig:fracs}
\end{figure}

For neutrino oscillation experiments such as those listed previously, the model of choice is the Bertini Cascade model. In this model, the incident hadron is injected into the nucleus and allowed to take a series of steps within the nuclear medium. At each step, the particle has a chance to interact with a particle within the nucleus and add it to the stack of active particles. These then step through the nuclear matter, possibly interacting along the way and activating more particles. This so-called cascade of particles continues until all active particles stop within or exit the nucleus and are added to the transport simulation outside of the nucleus. The latter are the interaction results that users can access and which Geant4Reweight uses to define the exculsive interaction channel. Geant4Reweight includes an application which produces effective exclusive cross sections from the Bertini Cascade that are used to calculate weights for exclusive interactions. It does this by invoking the Bertini Cascade model many times over a given momentum range to determine the fraction of times it ends in a specific final state. This fraction is then multiplied by the inelastic cross section to create exclusive cross sections. This is demonstrated in figure \ref{fig:fracs}, where the exclusive channel fractions are shown on the left and the exclusive cross sections are shown on the right along with the total inelastic cross section. These effective exclusive cross sections can be varied and reweighted in a similar way to the full inelastic cross section. 

The reweighting scheme is easily extended to include the exclusive cross sections. We treat the inelastic cross section as the sum over multiple exclusive channels:
\begin{equation}
    \sigma_I = \sum\limits_c^{N_c}\sigma_c.
\end{equation}
We allow users to specify a variation factor to each of the exclusive channels as well as to the total inelastic cross section similar to before:
\begin{equation}
    \sigma_I \rightarrow \sigma_I' = f_I\bigg(\sum\limits_c^{N_c}f_c\sigma_c\bigg).
\end{equation}
$W_S$, $W_E$, and $W_X$ remain largely the same with $\sigma'$ and $\sigma_{I}'$ changed accordingly. $W_I$, however, is modified to account for the separate variations to the exclusive channels. Here, we also have to account for the exclusive channel ($j$) of the interaction. For clarity, the subscript $i$ denoting that this occurs for a given step has been omitted.
\begin{equation}
    \begin{split}
    W_{I,j} &= \bigg(\frac{f_j \sigma_j}{\sum\limits_c^{N_c} f_c\sigma_c}\bigg)
                 \bigg(\frac{\sigma_I}{\sigma_j}\bigg)
                 \bigg(\frac{f_I\sum\limits_c^{N_c} f_c \sigma_c}{\sigma'}\bigg)
                 \bigg(\frac{\sigma}{\sigma_I}\bigg)
                 \bigg(\frac{1 - e^{-\Delta L\sigma'}}{1 - e^{-\Delta L\sigma}}\bigg)
    \\
          &= f_j f_I \bigg(\frac{\sigma}{\sigma'}\bigg)
             \bigg(\frac{1 - e^{-\Delta L\sigma'}}{1 - e^{-\Delta L\sigma}}\bigg)
    \end{split}
\end{equation}

\section{Reweighting examples}
\label{sec:example}
In this section, we present examples of validation of Geant4Reweight's weighting scheme applied to simulated particle tracks. Samples of $\pi^+$ were generated using a particle gun in the liquid argon of the ProtoDUNE-SP detector~\cite{protodune_tdr}\cite{protodune_performance} using the LArSoft toolkit~\cite{larsoft}. The samples had a Gaussian momentum profile centered at $1 ~\textrm{GeV/c}$ with a width of $100~\textrm{MeV/c}$. Within LArSoft's interface to Geant4, custom Geant4 physics lists were written to allow us to separately vary the inelastic and elastic $\pi^+$ cross sections. The generated hadrons were then propagated through liquid argon using the custom physics lists under three configurations:
\begin{enumerate}
    \item Nominal inelastic and elastic cross sections
    \item Inelastic cross section multiplied by 1.5. Nominal elastic cross section.
    \item Elastic cross section multiplied by 1.5. Nominal inelastic cross section.
\end{enumerate}
The results of this simulation were then passed to a LArSoft module that translated the LArSoft-style data products into Geant4Reweight objects. The nominal simulation results were then reweighted according to the varied configurations above.

\subsection{Changing the $\pi^+$ inelastic cross section}
\begin{figure}[htbp]
    \centering
    \begin{subfigure}[t]{0.45\textwidth}
        \centering
        \includegraphics[width=\textwidth]{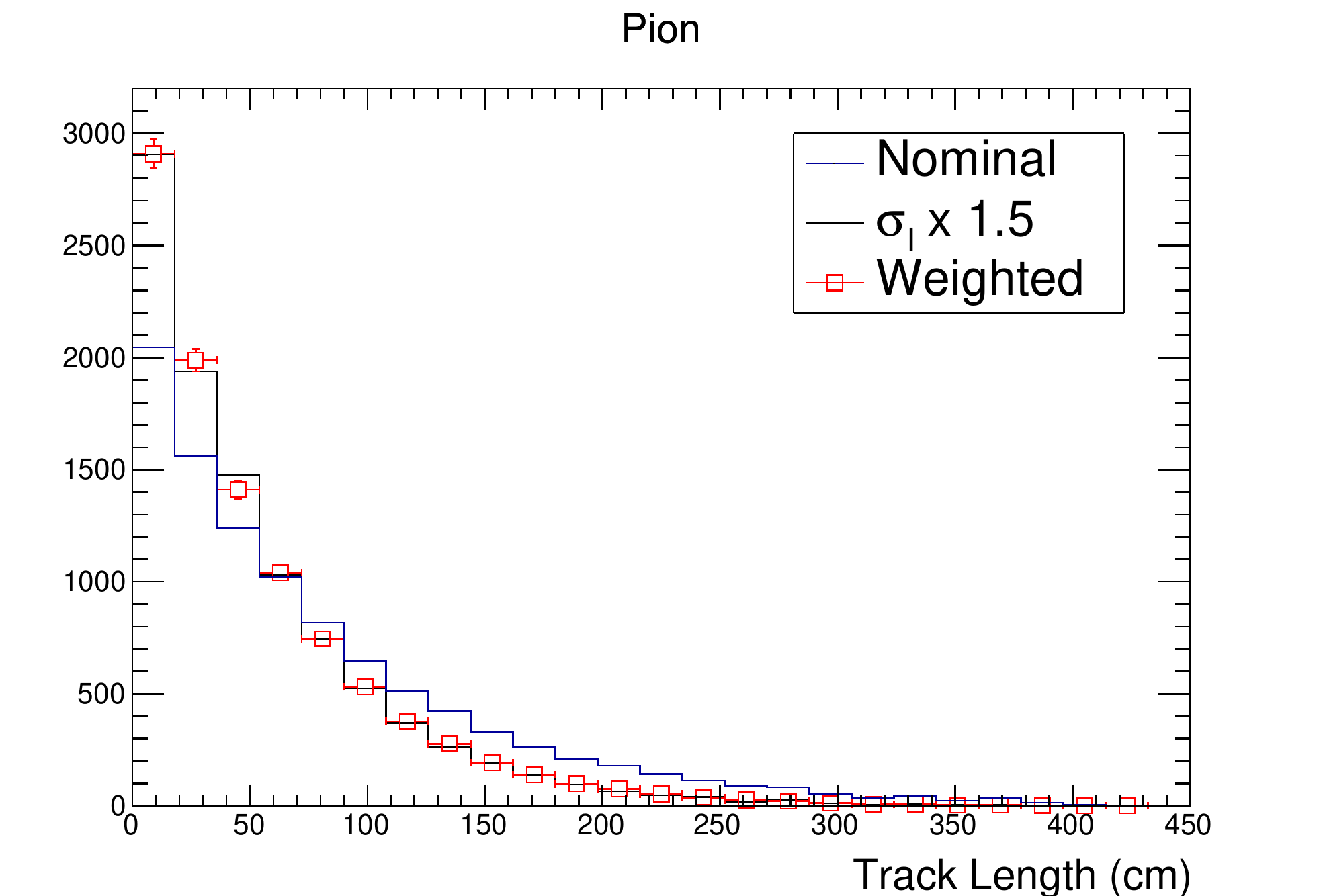}
        \caption{}
        \label{subfig:pion_inel_len}
    \end{subfigure}
    ~
    \begin{subfigure}[t]{0.45\textwidth}
        \centering
        \includegraphics[width=\textwidth]{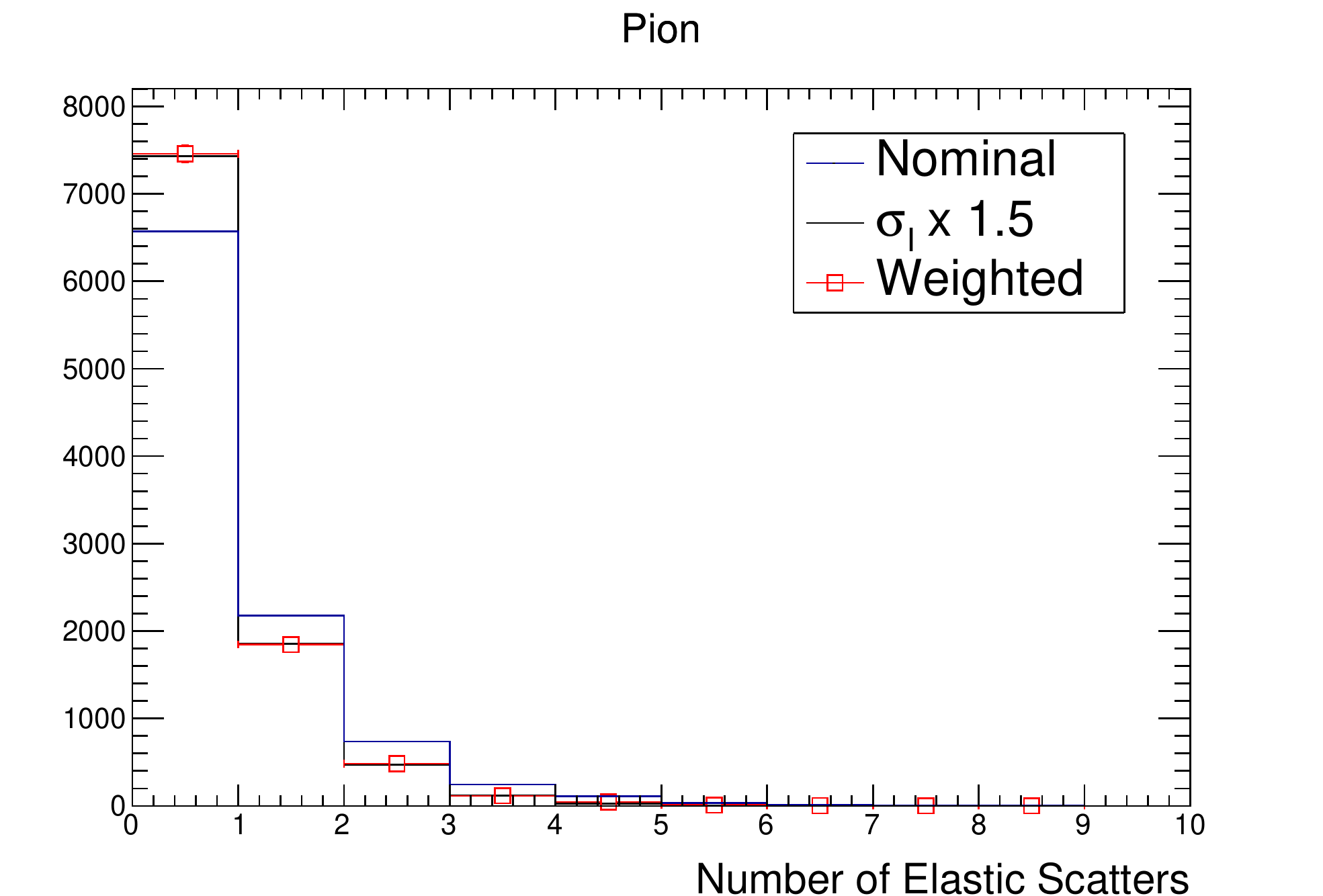}
        \caption{}
        \label{subfig:pion_inel_nElast}
    \end{subfigure}
    \caption{Comparison of pion track lengths (left) and number of elastic scatters (right) using Geant4 with nominal cross sections (Nominal) and with inelastic cross sections varied directly in Geant4 (Varied) and using Geant4Reweight (Weighted).  }
    \label{fig:pion_inel}
\end{figure}

Figure \ref{fig:pion_inel} shows comparison plots for the following samples: 1) running Geant4 with both of the nominal cross sections (``Nominal'' in the plots), 2) running Geant4 with the inelastic cross section multiplied by 1.5 (``Varied''), and 3) applying Geant4Reweight to the nominal cross section sample according to the inelastic cross section change (``Weighted''). In figure \ref{subfig:pion_inel_len}, the track length of the particle is shown. This is the integrated path length traveled by the $\pi^+$ from its creation to the point at which it stops, decays in flight, or undergoes an inelastic interaction. Increasing the inelastic cross section increases the chance that it will interact earlier, before stopping or decaying, thus shifting the Weighted and Varied distributions lower, as expected. Figure \ref{subfig:pion_inel_nElast} shows the number of elastic scatters performed by the $\pi^+$ during its travel. Similarly, increasing the inelastic cross sections means the $\pi^+$ will be less likely to travel as far as in the nominal simulation, suppressing the chance for elastic scatters to occur. The figures show good agreement between the varied and weighted samples, supporting the validity of our reweighting scheme. 

\subsection{Changing the $\pi^+$ elastic cross section}
\begin{figure}[htbp]
    \centering
    \begin{subfigure}[t]{0.45\textwidth}
        \centering
        \includegraphics[width=\textwidth]{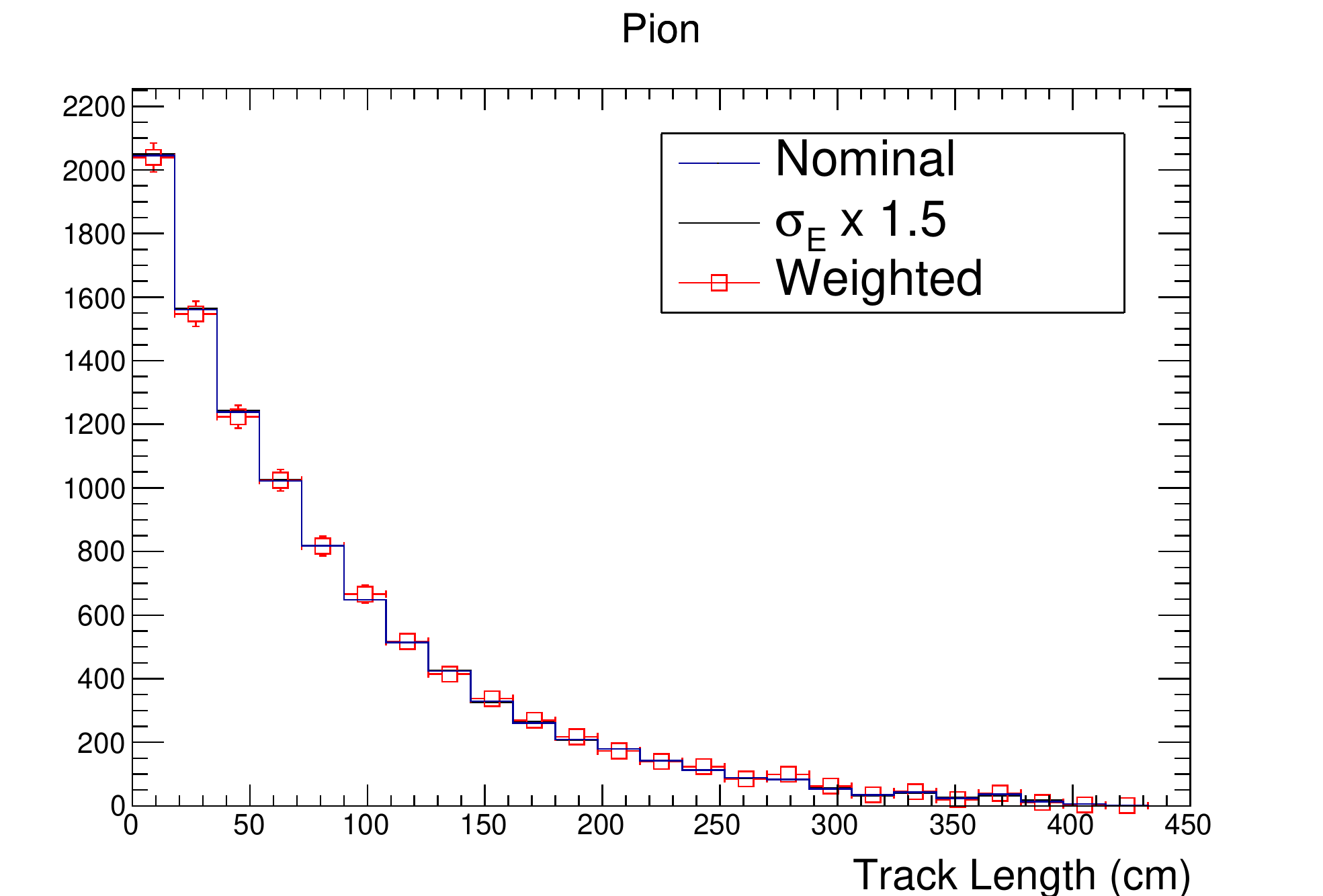}
        \caption{}
        \label{subfig:pion_el_len}
    \end{subfigure}
    ~
    \begin{subfigure}[t]{0.45\textwidth}
        \centering
        \includegraphics[width=\textwidth]{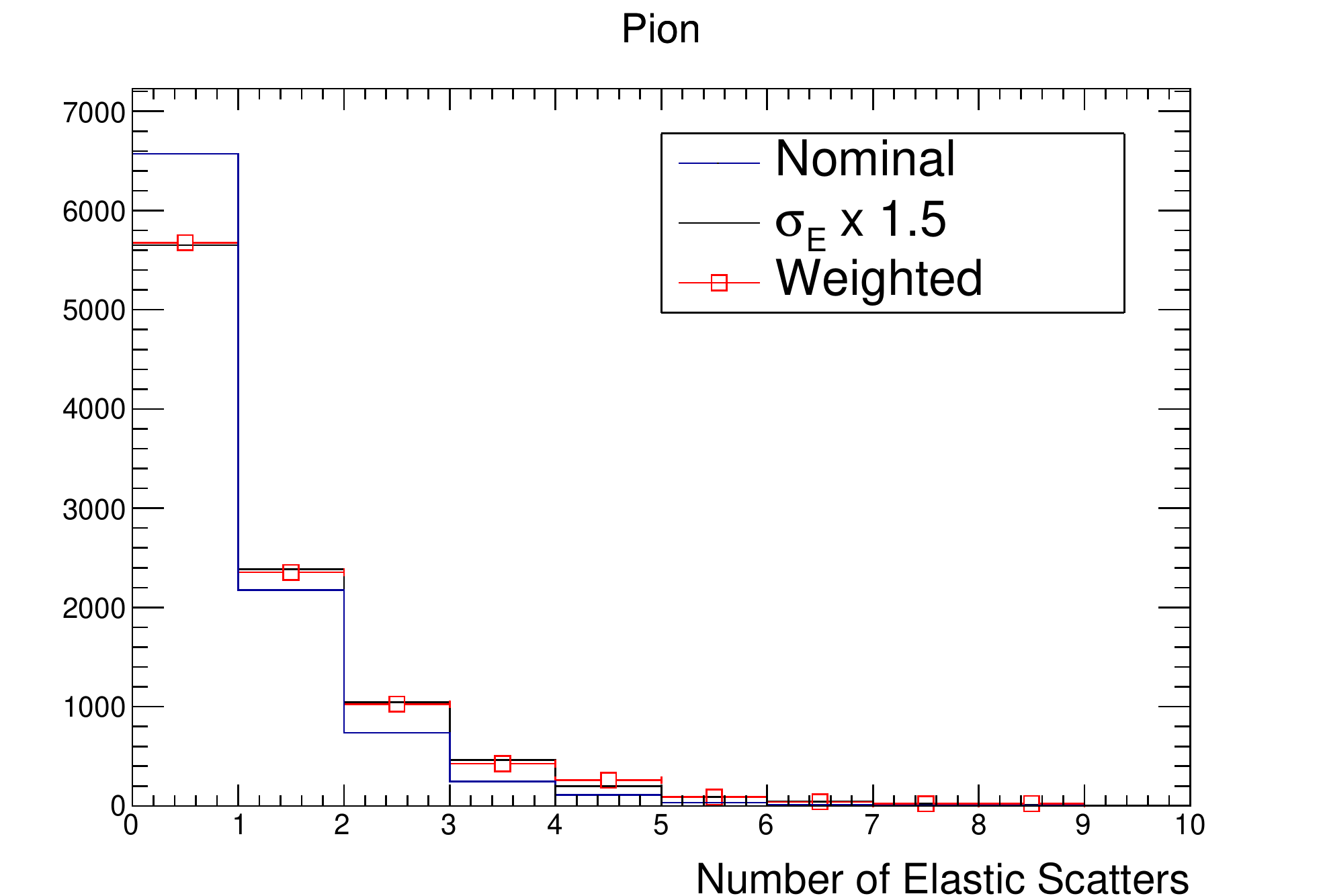}
        \caption{}
        \label{subfig:pion_el_nElast}
    \end{subfigure}
    \caption{Comparison of pion track lengths (left) and number of elastic scatters (right) using Geant4 with nominal cross sections (Nominal) and with elastic cross sections varied directly in Geant4 (Varied) and using Geant4Reweight (Weighted).  }
    \label{fig:pion_el}
\end{figure}

Figure \ref{fig:pion_el} shows similar comparison plots as the previous section, but for the sample with the elastic cross section multiplied by 1.5 and the inelastic cross section kept the same. In figure \ref{subfig:pion_el_len}, the track length distribution shows no significant change. This is due to the track-length definition above, as well as the fact that the elastic interaction in Geant4 removes minimal energy from the $\pi^+$. As such, more elastic scatters do not cause the $\pi^+$ to stop earlier. Figure \ref{subfig:pion_el_nElast} shows that the number of elastic scatters performed by the $\pi^+$ increases as expected.

\subsection{Recovering the nominal cross section}
\begin{figure}[htbp]
    \centering
    \begin{subfigure}[t]{0.45\textwidth}
        \centering
        \includegraphics[width=\textwidth]{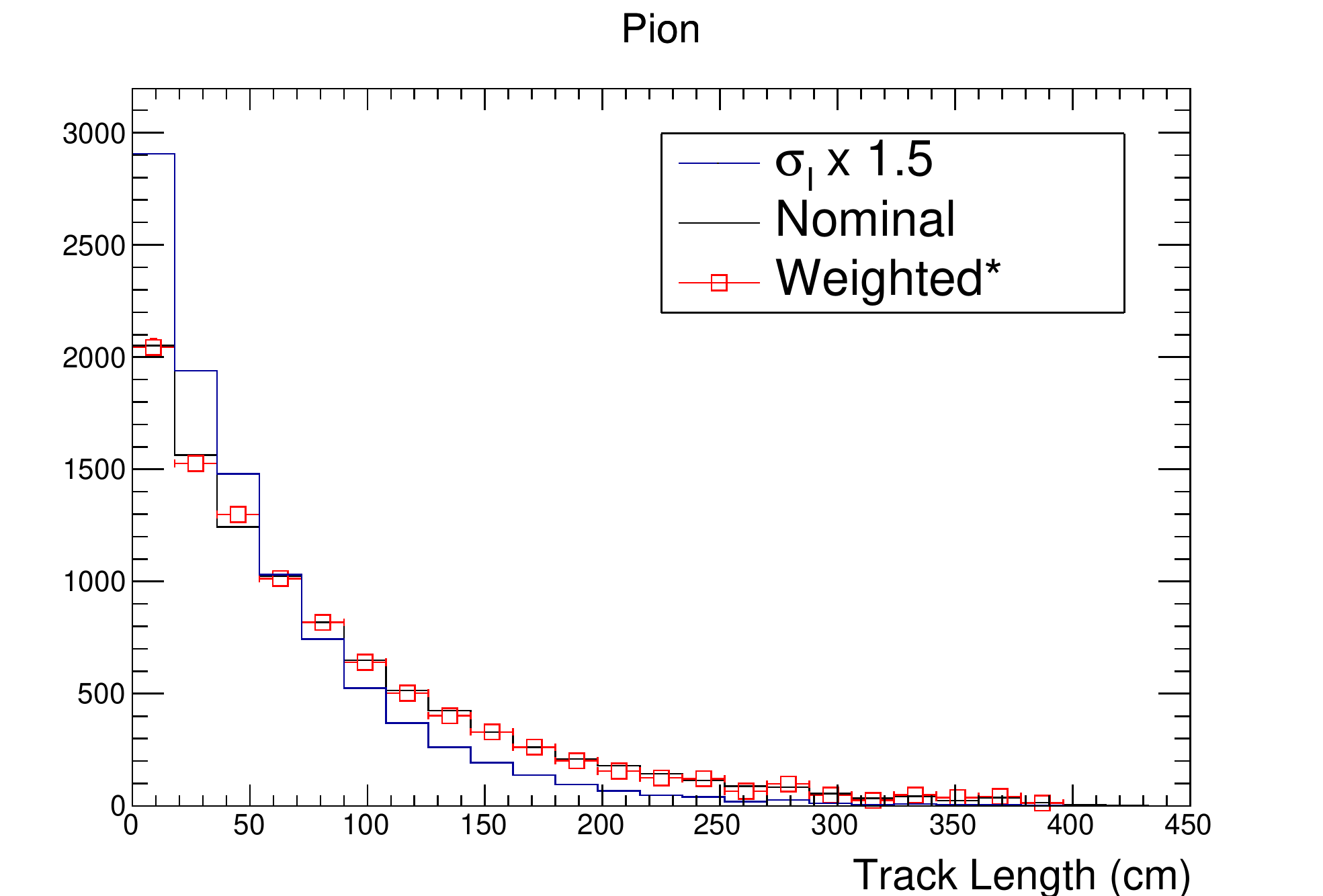}        
        \caption{}
        \label{subfig:pion_inel_len_2}
    \end{subfigure}
    ~
    \begin{subfigure}[t]{0.45\textwidth}
        \centering
        \includegraphics[width=\textwidth]{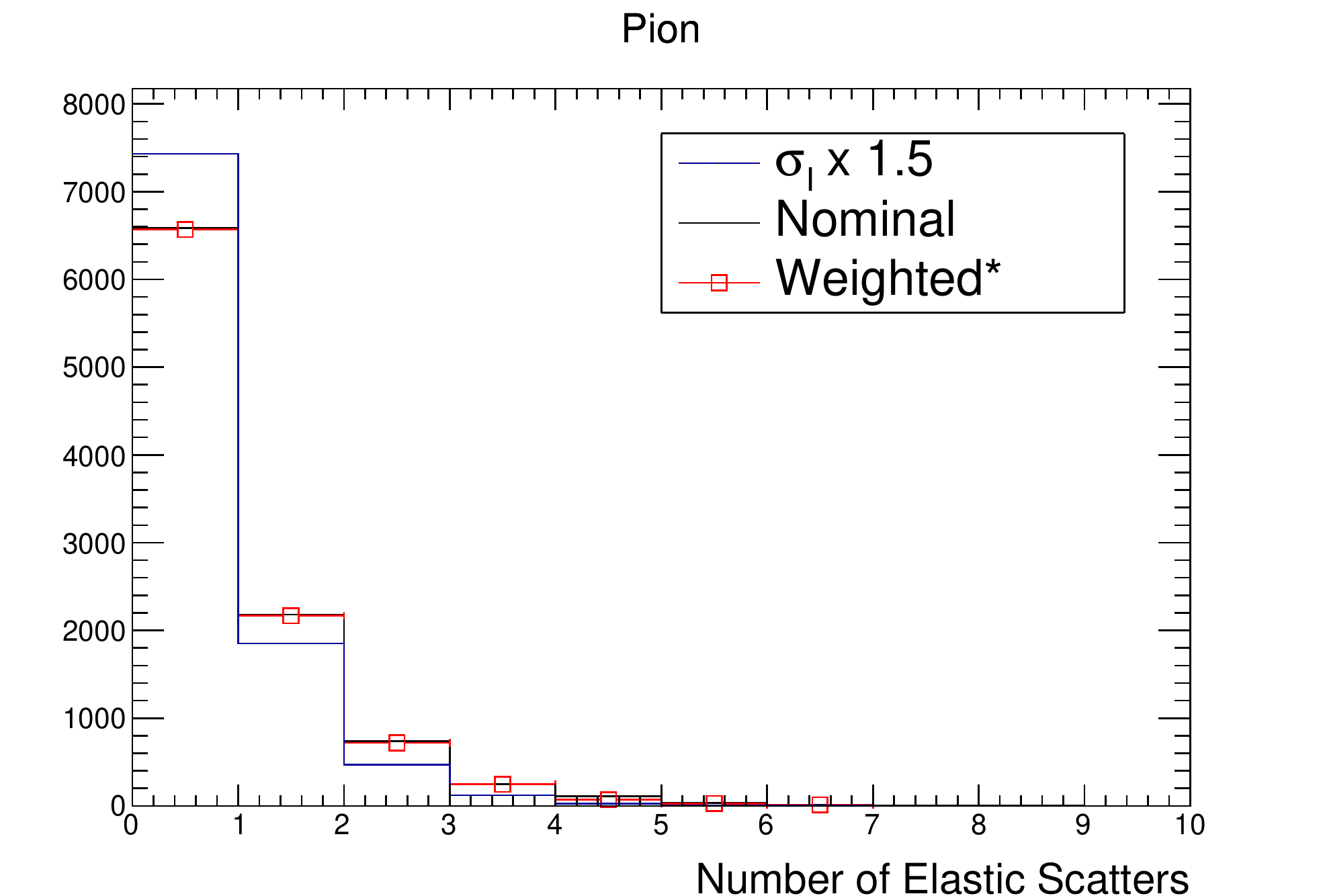}
        \caption{}
        \label{subfig:pion_inel_nElast_2}
    \end{subfigure}
    \caption{Comparison of pion track lengths (left) and number of elastic scatters (right) using Geant4 with nominal cross sections (Nominal) and with inelastic cross sections varied directly in Geant4 (Varied) and using Geant4Reweight applied to the Varied sample  (Weighted*).}
    \label{fig:pion_inel_2}
\end{figure}

As a final demonstration, we show the ability for Geant4Reweight to recover the nominal cross sections in Geant4 by reweighting the samples generated with varied cross sections. This is shown first in figure~\ref{fig:pion_inel_2}. Here, ``Nominal'' and ``Varied'' are samples generated with nominal cross sections and with the inelastic cross section scaled by 1.5. ``Weighted*'' shows a reweighting applied to the Varied sample, where a scaling factor of 2/3 is applied to the varied cross section to return to the nominal. The figure shows the ability to reweight from a varied cross section back to nominal.

\begin{figure}[htbp]
    \centering
    \begin{subfigure}[t]{0.45\textwidth}
        \centering
        \includegraphics[width=\textwidth]{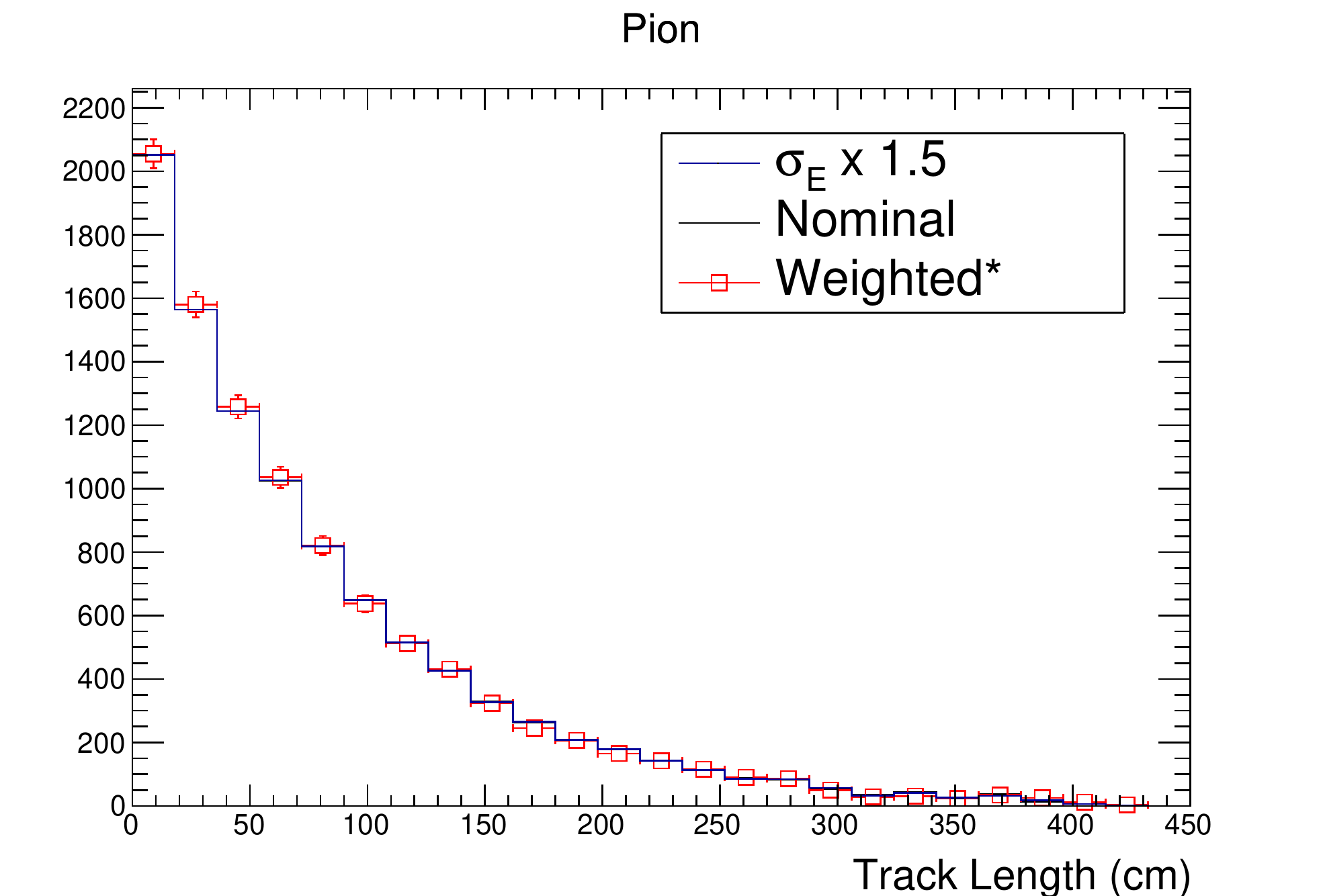}
        \caption{}
        \label{subfig:pion_el_len_2}
    \end{subfigure}
    ~
    \begin{subfigure}[t]{0.45\textwidth}
        \centering
        \includegraphics[width=\textwidth]{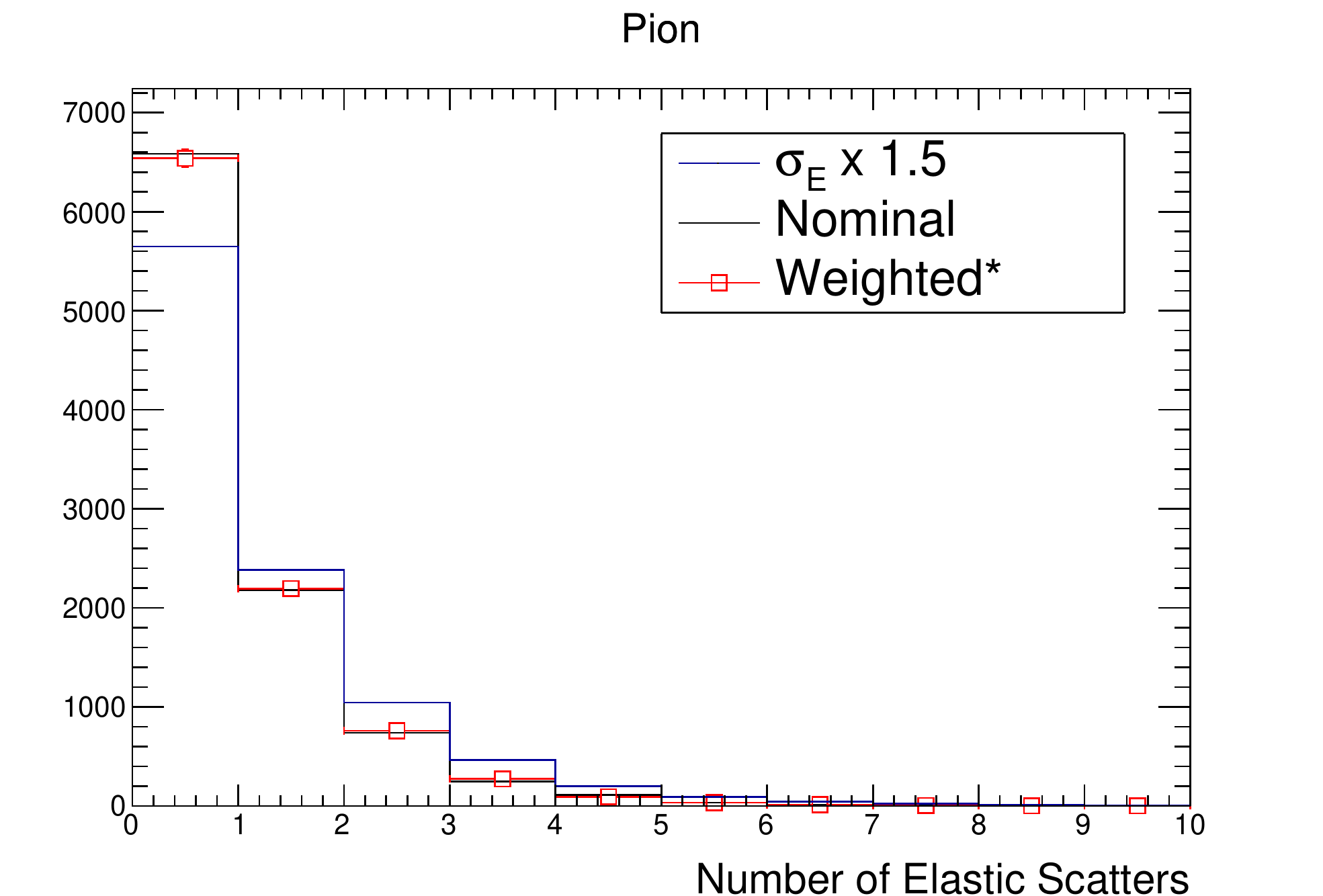}
        \caption{}
        \label{subfig:pion_el_nElast_2}
    \end{subfigure}
    \caption{Comparison of pion track lengths (left) and number of elastic scatters (right) using Geant4 with nominal cross sections (Nominal) and with elastic cross sections varied directly in Geant4 (Varied) and using Geant4Reweight applied to the Varied sample  (Weighted*).}
    \label{fig:pion_el_2}
\end{figure}

Figure \ref{fig:pion_el_2} shows a similar treatment, but with the elastic cross section increased by 50\% and then weighted to recover the nominal model. The nominal elastic cross section can be recovered via reweighting as shown.

\section{Fitting framework}
\label{sec:fitting}
Geant4Reweight includes a fitting framework to estimate uncertainties on the Geant4 cross sections. These uncertainties can then be propagated through physics analysis via reweighting. Sets of data are provided by Geant4Reweight, and users are able to include additional data as well. This framework uses the MIGRAD algorithm from the MINUIT package implemented within ROOT. Users provide a configuration file defining the following:
\begin{enumerate}
    \item The input final state fraction files. These are used as in section~\ref{subsec:exc_chan} to define exclusive cross sections.
    \item The data sets against which the Geant4 predictions are fit. These include the following:
        \begin{enumerate}
            \item  The path to the ROOT file containing the data points
            \item The incident particle type and target material.
            \item The category of cross sections within the data file (i.e. Total Inelastic, Absorption, etc.).
            \item The name of the ROOT objects representing the cross section data points within the data file.
        \end{enumerate}
    \item A set of parameters which represent variations to the cross sections across defined regions of momentum.
    \item Configurations for the fit minimizer. 
\end{enumerate}

\subsection{Parameter definitions}
The parameters used within the fit are user-defined variations across regions of incident hadron momentum, for the various cross sections. For charged pions, this includes the exclusive cross sections defined in table~\ref{tab:pi_exc}, as well as the total inelastic (the sum of the exclusive channels --- also referred to as reaction) and elastic cross sections. The reaction cross section parameters work simultaneously with the exclusive parameters, similar to the weighting scheme in section~\ref{sec:reweighting}. Flat variations are applied across the momentum range chosen by users. 

During each iteration of the fit, the values of the user-defined parameters are changed. The cross sections are scaled according to the variations defined by the new parameter values, and fit to the provided data. If the fit is successful, the best-fit values of the parameters and their covariances are saved in the output ROOT file.


\subsection{$\chi^2$ definitions}

The data in the fit is categorized by the interaction channel of each measurement. The relevant channels are the exclusive channels (such as those listed in table \ref{tab:pi_exc}), combinations of these (such as pion absorption + charge exchange), the elastic interaction, the reaction (total inelastic) interaction, and the inclusive interaction (reaction + elastic). The quantity of measurements for each exclusive channel varies widely, and a simple $\chi^2$ calculation will likely lead to channels with more measurements dominating the fit. Instead, a reduced $\chi^2$ statistic is calculated for each channel, as shown in equation \ref{eq:exc_chi2}. Here, $k$ denotes the interaction channel, $N_k$ is the number of cross section measurements for channel $k$, and $p_k$ is the number of user-defined parameters for that channel. $\sigma^{\textrm{MC}}$ denotes the Geant4 cross section that has been varied by the fit parameters as described above.


\begin{align}
\chi^2_k &= \frac{1}{N_k - p_k} \sum_{i}^{N_k} \left(\frac{\sigma^{\textrm{
Data}}_i - \sigma^{\textrm{MC}}_i}{\Delta\sigma^{\textrm{Data}}_i} \right)^2\label{eq:exc_chi2} \\
\chi^2 &= \frac{1}{N_{\textrm{C}} + f - 1}\bigg(\sum\limits_{k} \chi^2_k + f\chi^{2}_{\textrm{Total}}\bigg)  \label{eq:total_chi2}
\end{align}
The full $\chi^2$ is shown in equation \ref{eq:total_chi2}. Here, the $\chi^2$ contributions from all channels other than the inclusive interaction are simply summed. The inclusive interaction is treated slightly differently because variations to the total cross section can sometimes dominate the fit. As such, a scaling parameter $f$ is introduced, allowing the user to control the contribution of the inclusive channel to the full $\chi^2$. The leading factor is the effective number of degrees of freedom, where $N_C$ is the number of channels included in the sum. This term ensures $\chi^2 \sim 1$ as $\big(\sigma^{\textrm{
Data}}_i - \sigma^{\textrm{MC}}_i\big)$ approaches  $\Delta\sigma^{\textrm{Data}}$.

\subsubsection{Fit with correlated data}
The reduced $\chi^2$ statistic defined in equation \ref{eq:exc_chi2} is valid for uncorrelated data. 
However, a recent measurement by the DUET collaboration includes correlations in their published results for $\pi^+$ absorption and charge exchange on carbon~\cite{DUET}. We have included this data in the fit in a specific way to account for these correlations. A standard $\chi^2$ definition is shown in equation \ref{eq:DUET_chi2}. For this set of data, the absorption and charge exchange cross sections were each measured at five momenta. As such, the sum in this equation runs over all ten data points. In order to treat the measurements of these two channels in a similar manner as above, we have chosen to split this value equally between the modified reduced $\chi^2$ statistics for the absorption and charge exchange channels, as shown in equations \ref{eq:abs_chi2_special} and \ref{eq:cex_chi2_special}. The denominator of the first term in each of these equations includes the five data points for each channel. 

\begin{equation}\label{eq:DUET_chi2}
    \chi^2_{\textrm{DUET}} = \sum\limits_{i,j}^{10}\big(\sigma^{\textrm{DUET}}_i - \sigma^{\textrm{MC}}_i\big)V_{ij}^{-1} \big(\sigma^{\textrm{DUET}}_j - \sigma^{\textrm{MC}}_j\big)
\end{equation}

\begin{equation}\label{eq:abs_chi2_special}
    \chi^2_{\textrm{Abs}} = \frac{1}{N_{\textrm{Abs}} + 5 - p_{\textrm{Abs}}}\Bigg(\sum\limits^{N_{\textrm{Abs}}}_{i}\bigg(\frac{\sigma_i^{\textrm{Data}} - \sigma_i^{\textrm{MC}}}{\Delta\sigma^{\textrm{Data}}_i}\bigg)^2 + \frac{\chi^2_{\textrm{DUET}}}{2}\Bigg)
\end{equation}

\begin{equation}\label{eq:cex_chi2_special}
    \chi^2_{\textrm{Cex}} = \frac{1}{N_{\textrm{Cex}} + 5 - p_{\textrm{Cex}}}\Bigg(\sum\limits^{N_{\textrm{Cex}}}_{i}\bigg(\frac{\sigma_i^{\textrm{Data}} - \sigma_i^{\textrm{MC}}}{\Delta\sigma^{\textrm{Data}}_i}\bigg)^2 + \frac{\chi^2_{\textrm{DUET}}}{2}\Bigg)
\end{equation}
In general, other correlated data can be included in the fit in a similar manner. If such a set of data exists, a user can implement their own calculation of its contribution to the specific $\chi^2_k$.


\subsection{Example fit}For demonstration, we have fit Geant4 to a collection of $\pi^+$-C data. The parameters we have chosen to use are shown in table \ref{tab:parameters}. The data included in the fit is shown in table \ref{tab:fit_data}. Throughout this section, the ``reaction'' channel is used as another term for any inelastic pion-nucleus interaction. Its cross section is the combination of all exclusive channels (i.e. it is the total inelastic cross section shown in figure \ref{fig:xsec}).

\begin{table}[htbp]
\centering
\caption{\label{tab:parameters} The parameters used within the demonstration fit.}
\smallskip
    \begin{tabular}{|l|l|l|}
    \hline
    Parameter Name & Channel & Momentum Range [MeV$/c$]\\
    \hline
    fReacLow & Reaction & 10--200 \\
    fReacHigh & Reaction & 700--2005 \\
    fAbs & Absorption & 200--700 \\
    fCex & Charge Exchange & 200--700 \\
    fInel & Quasielastic & 200--700 \\
    \hline
    \end{tabular}
\end{table}

\begin{table}[htbp]
\centering
\caption{\label{tab:fit_data} The data sets used within the demonstration fit.}
\smallskip
    \begin{tabular}{|l|l|l|}
    \hline
    Reference & Channel(s) & Momentum Range [MeV/$c$]\\
    \hline
    B. W. Allardyce \textit{et al.}~\cite{Allardyce} & Reaction & 710--2000\\
    D. Ashery \textit{et al.}~\cite{Ashery2173} & Quasielastic, Abs. + Ch. Ex. & 175--433\\
    D. Ashery \textit{et al.}~\cite{Ashery946} & Charge Exchange & 265\\
    E. Bellotti \textit{et al.}~\cite{Bellotti} & Absorption, Charge Exchange & 230\\
    S. M. Levenson \textit{et al.}~\cite{Levenson} & Quasielastic & 194--417\\
    O. Meirav \textit{et al.}~\cite{Meirav} & Reaction & 128--170\\
    I. Navon \textit{et al.}~\cite{Navon} & Abs. + Ch. Ex. & 128\\
    E. S. Pinzon-Guerra \textit{et al.}~\cite{DUET} & Absorption, Charge Exchange & 206--295 \\
    A. Saunders \textit{et al.}~\cite{Saunders} & Reaction & 116--150\\
    \hline
    \end{tabular}
\end{table}

\begin{figure}[htbp]
    \centering
    \begin{subfigure}[t]{0.4\textwidth}
        \centering
        \includegraphics[width=\textwidth]{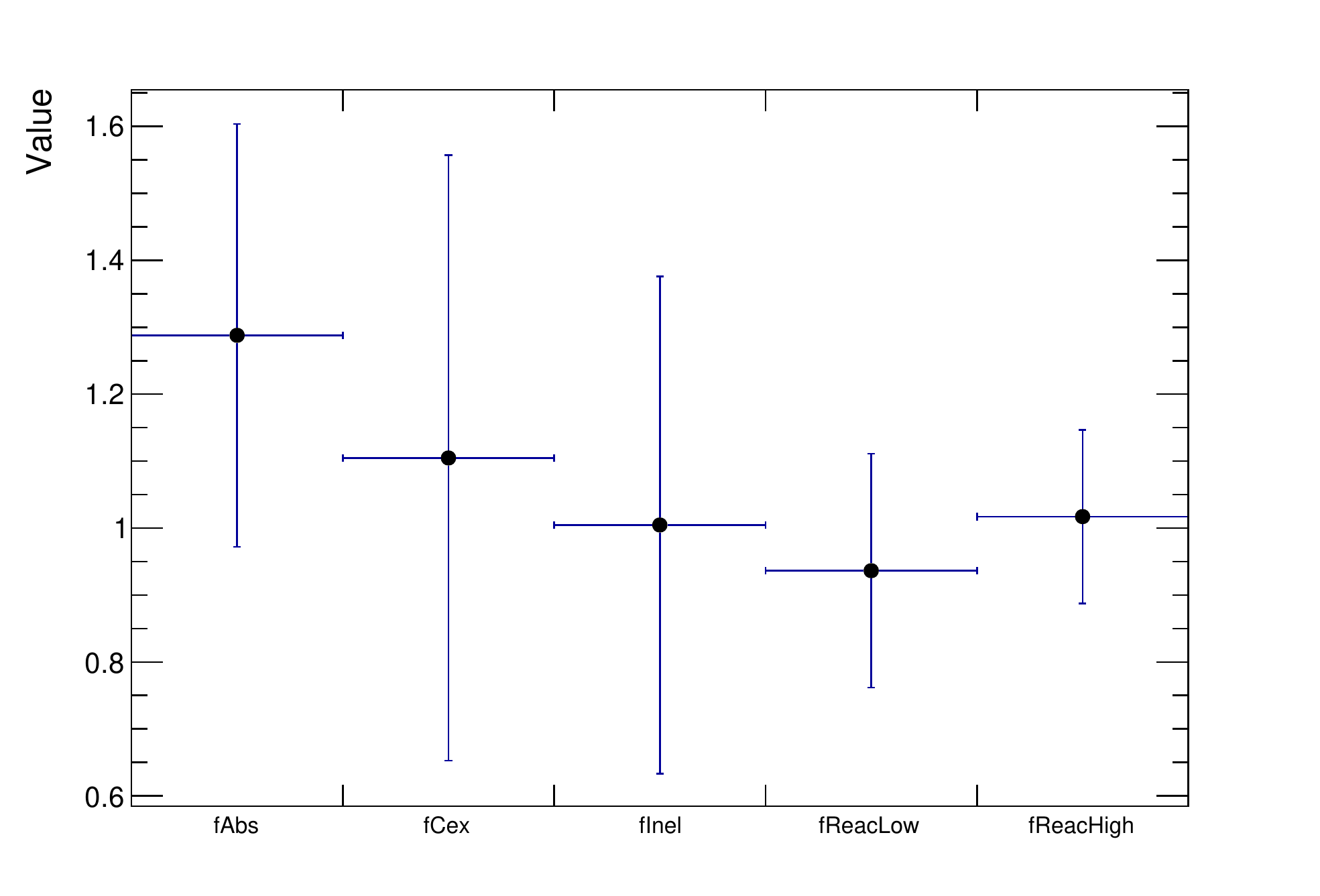}
        \caption{Best-fit parameter values and 1-D uncertainty.}
    \end{subfigure}
    ~
    \begin{subfigure}[t]{0.4\textwidth}
        \centering
        \includegraphics[width=\textwidth]{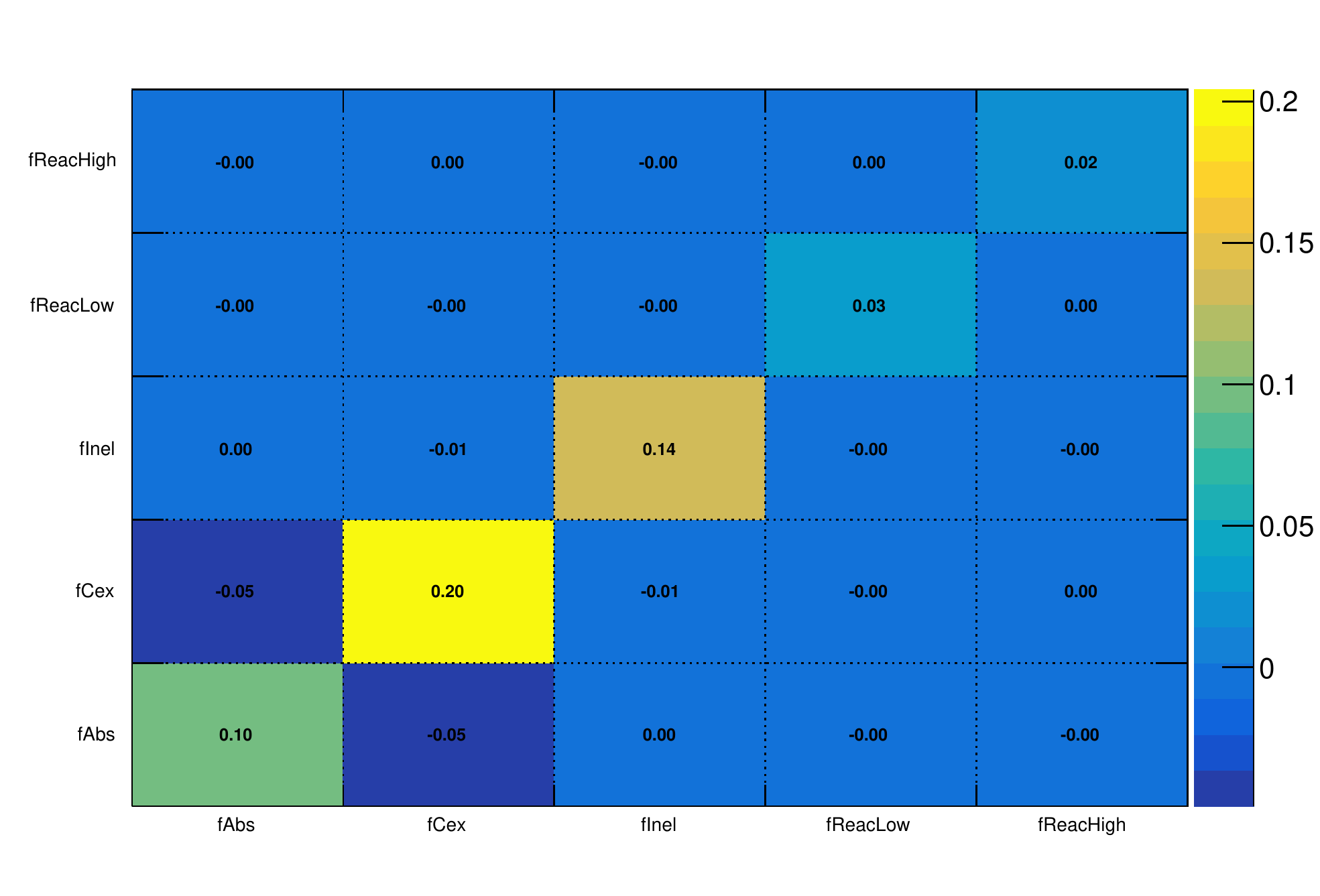}
        \caption{Output covariance matrix.}
    \end{subfigure}
    \caption{}
    \label{fig:pars_cov}
\end{figure}

\begin{figure}[!htbp]
    \centering
    \begin{subfigure}[t]{0.4\textwidth}
        \centering
        \includegraphics[width=\textwidth]{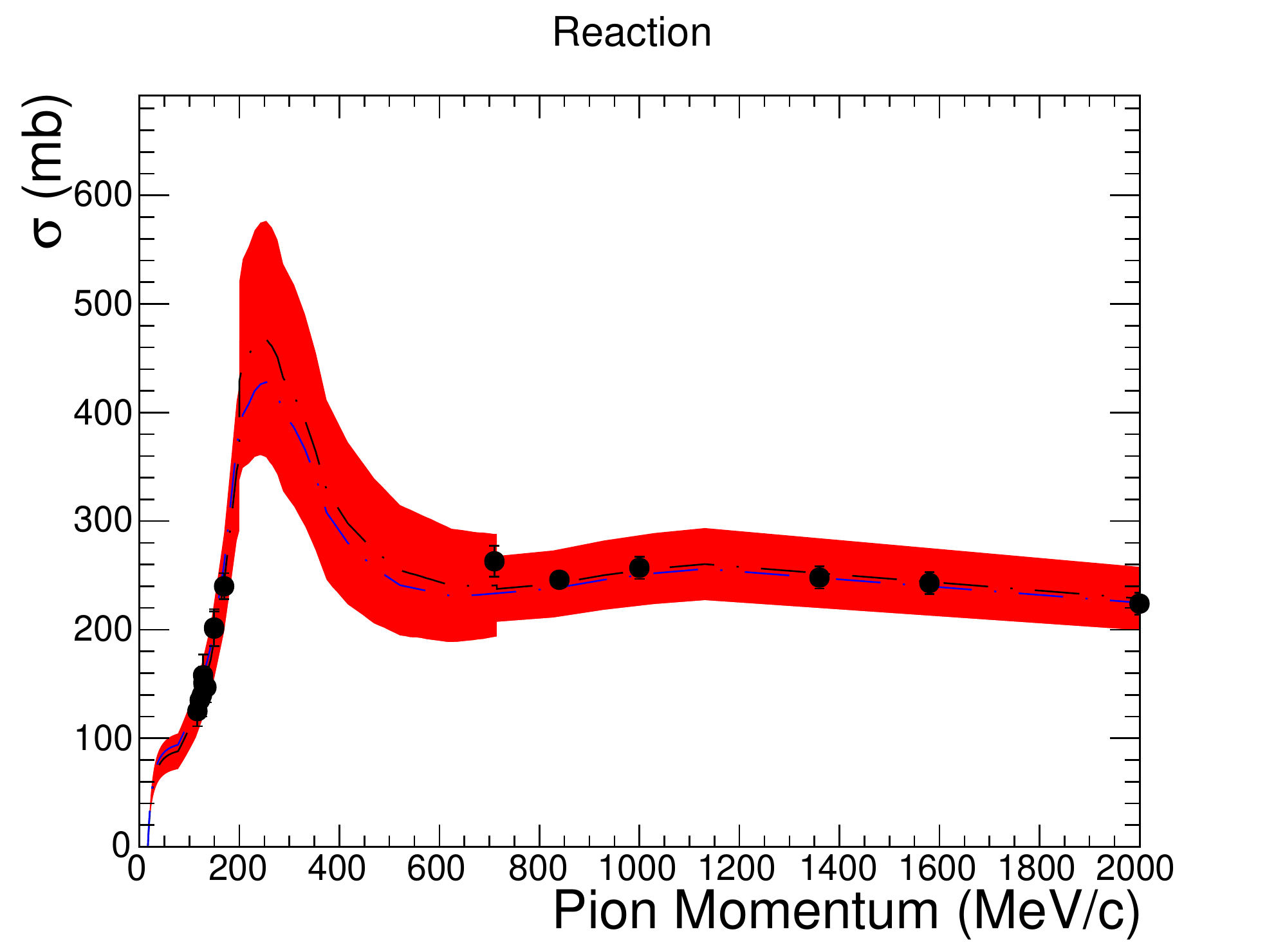}
        \caption{}
    \end{subfigure}
    ~
    \begin{subfigure}[t]{0.4\textwidth}
        \centering
        \includegraphics[width=\textwidth]{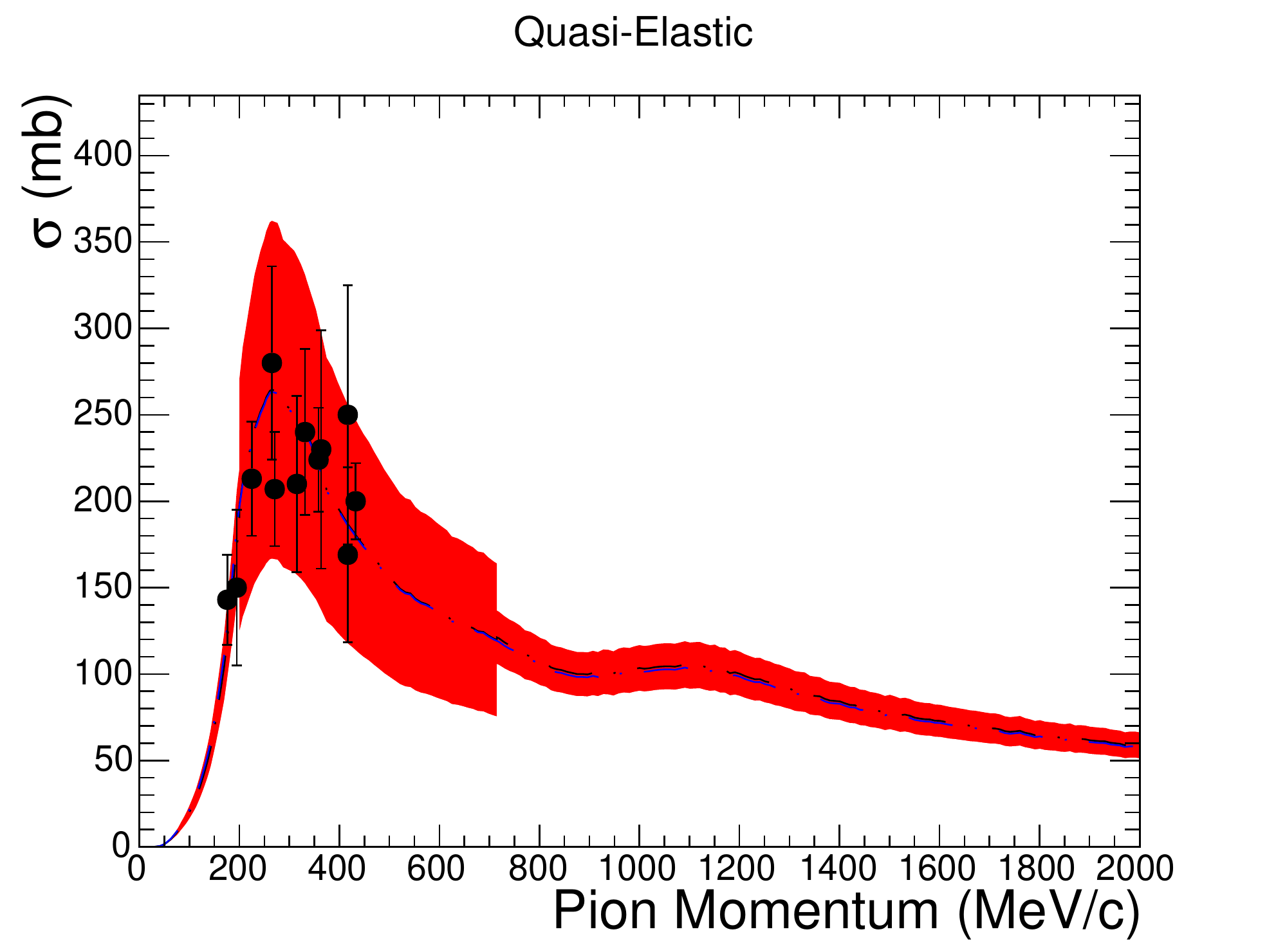}
        \caption{}
    \end{subfigure}
    
        \begin{subfigure}[t]{0.4\textwidth}
        \centering
        \includegraphics[width=\textwidth]{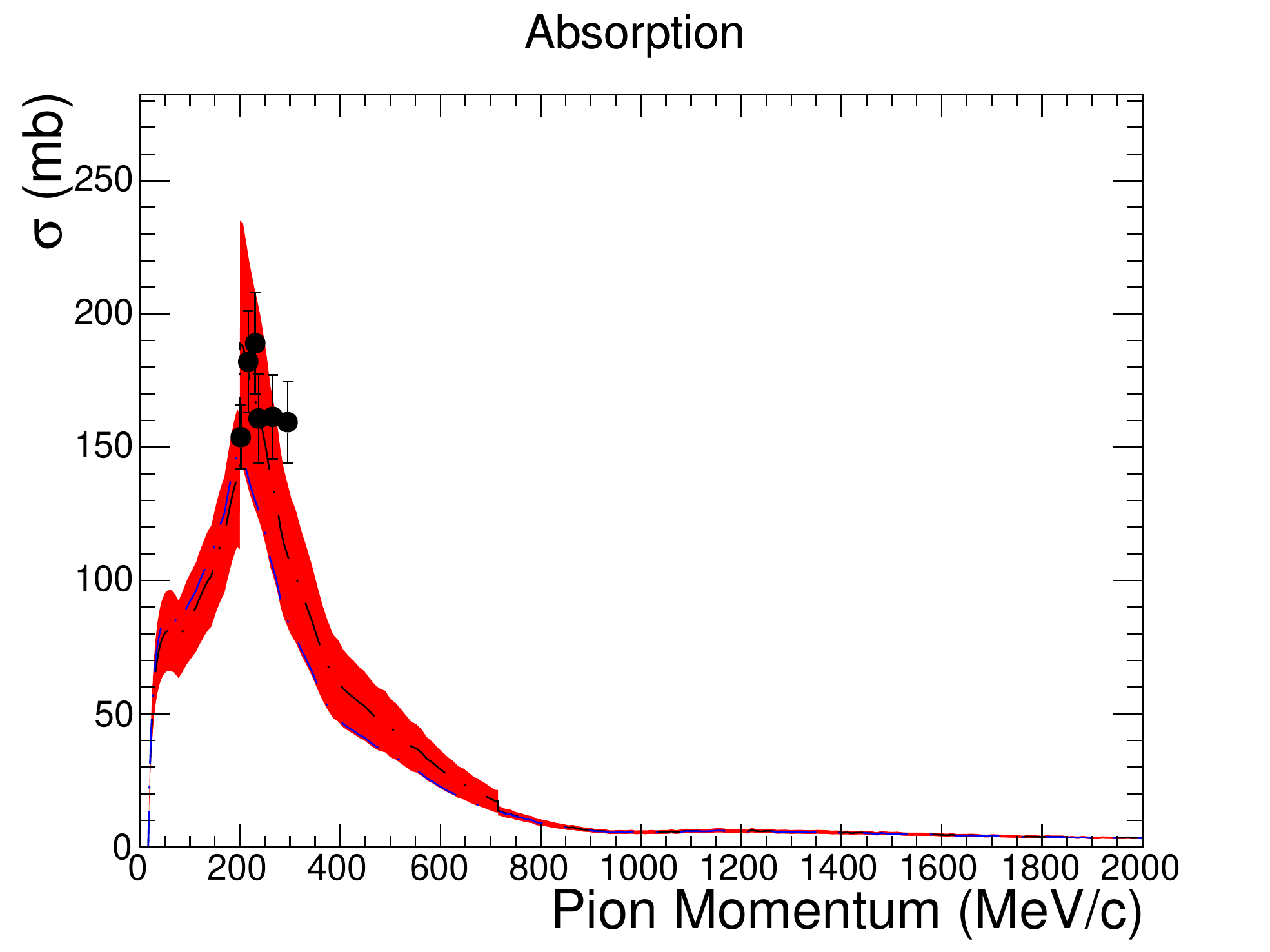}
        \caption{}
    \end{subfigure}
    ~
    \begin{subfigure}[t]{0.4\textwidth}
        \centering
        \includegraphics[width=\textwidth]{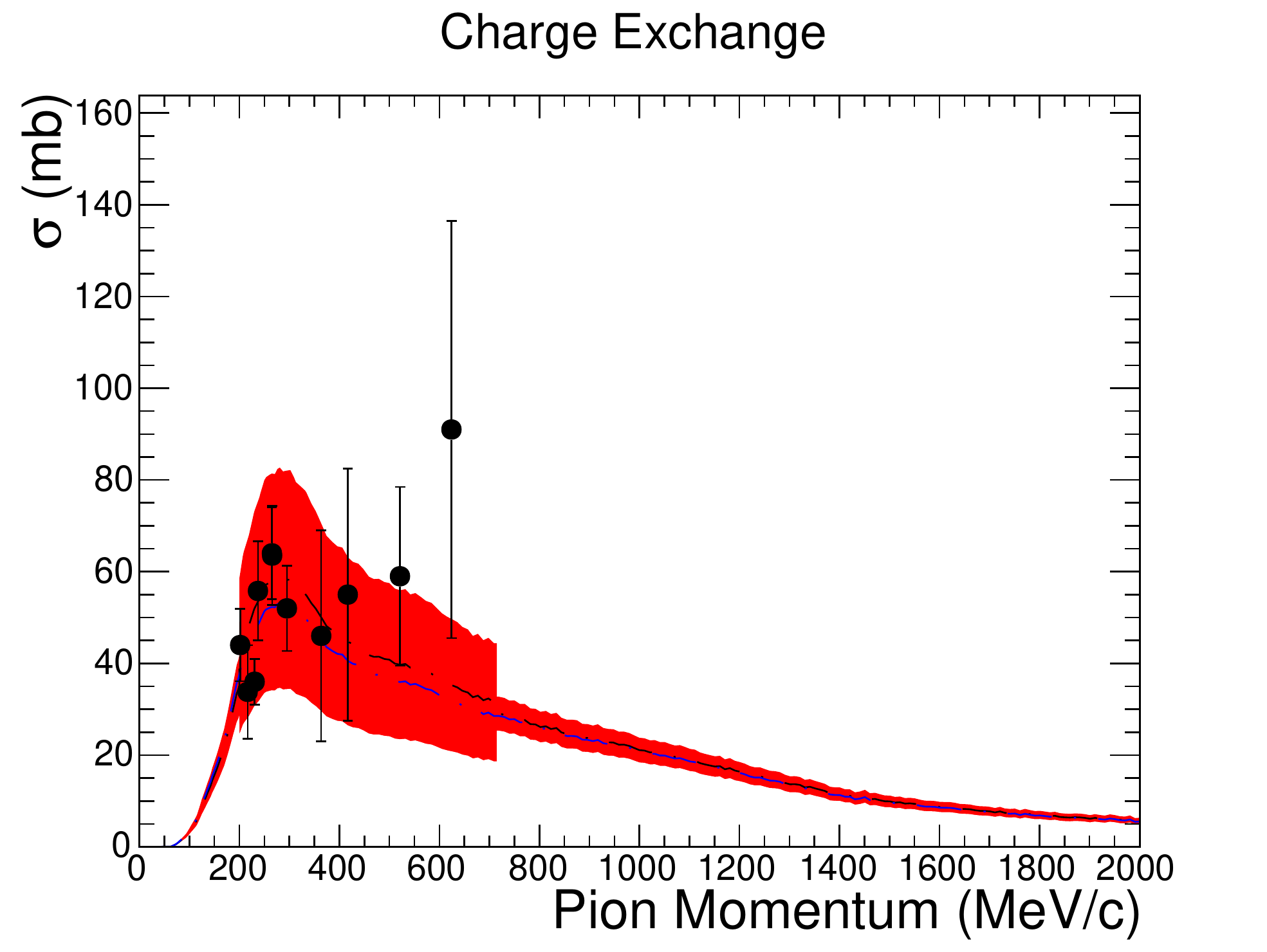}
        \caption{}
    \end{subfigure}
    
    \begin{subfigure}[t]{0.4\textwidth}
        \centering
        \includegraphics[width=\textwidth]{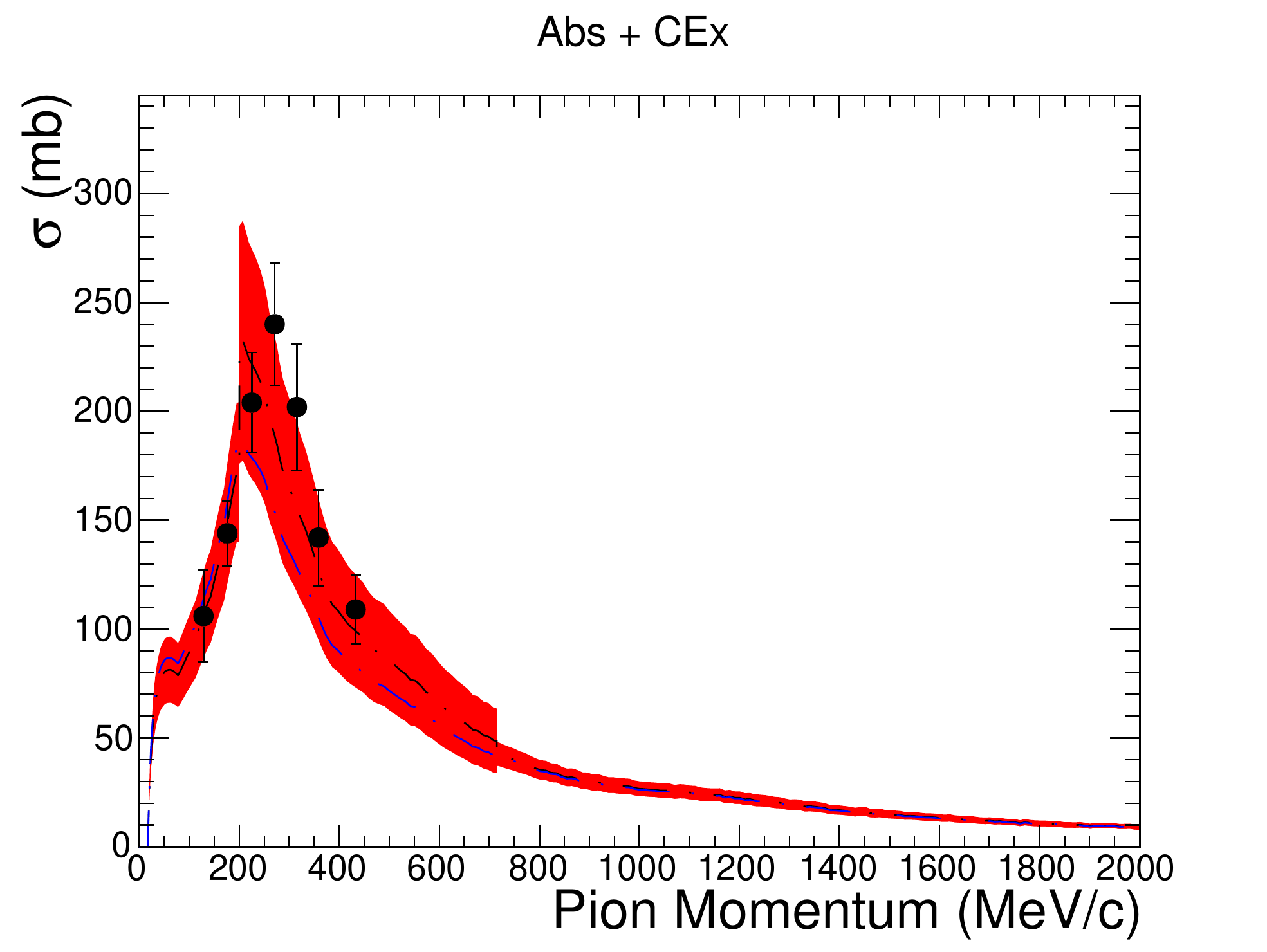}
        \caption{}
    \end{subfigure}
    \caption{Fit results of $\pi^+$-C cross sections extracted from Geant4 (with the Bertini Cascade model) fit to data (listed in table \ref{tab:fit_data}). The parameters used in the fit are defined in table \ref{tab:parameters}. The blue dashed line is the nominal cross section when extracted from Geant4, the black dashed line is the best fit value, and the red bands are calculated by setting the parameters to their $\pm1\sigma$ values.}
    \label{fig:fit}
\end{figure}

\begin{figure}[!htbp]
    \centering
    \begin{subfigure}[t]{0.4\textwidth}
        \centering
        \includegraphics[width=\textwidth]{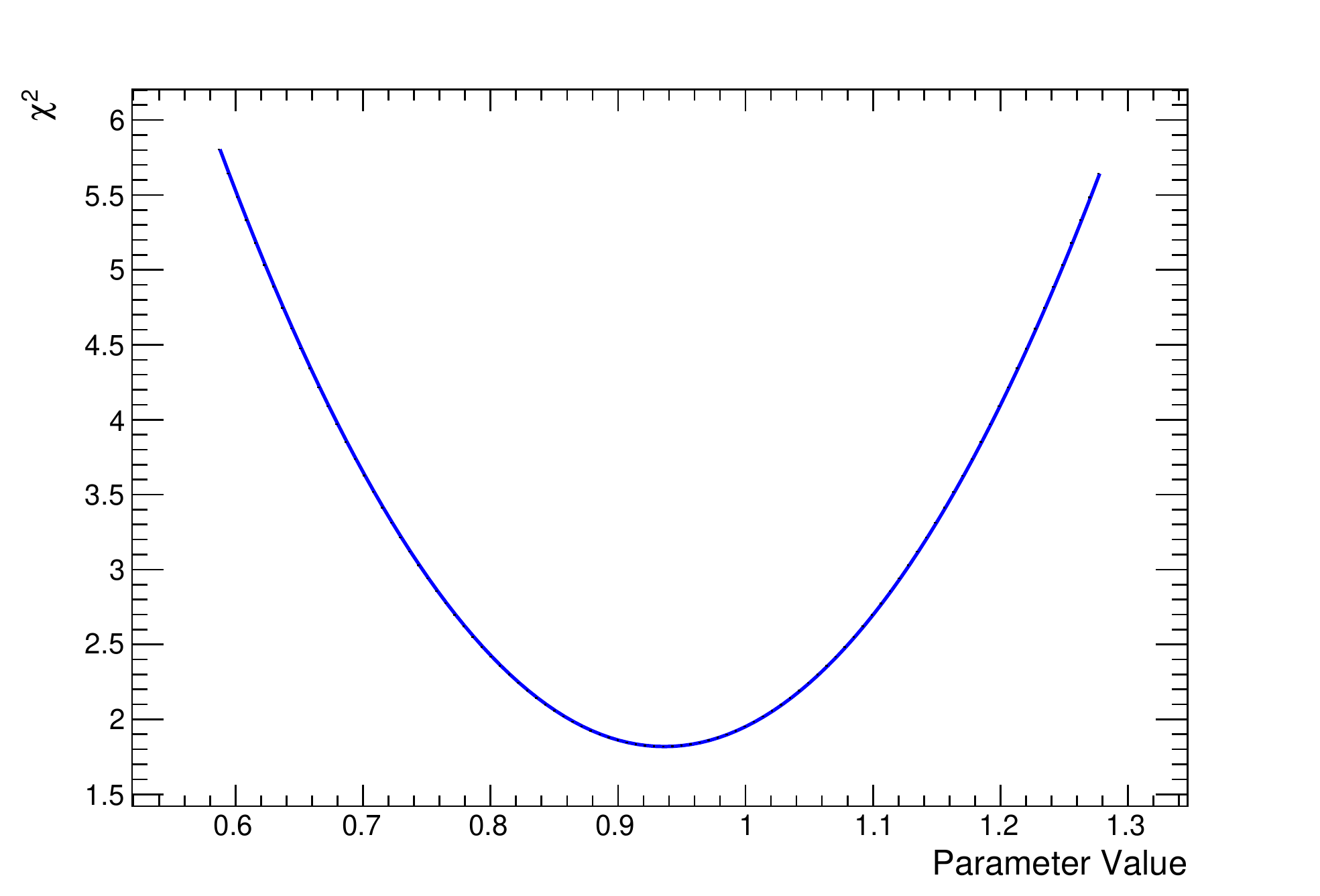}
        \caption{fReacLow}
    \end{subfigure}
    ~
    \begin{subfigure}[t]{0.4\textwidth}
        \centering
        \includegraphics[width=\textwidth]{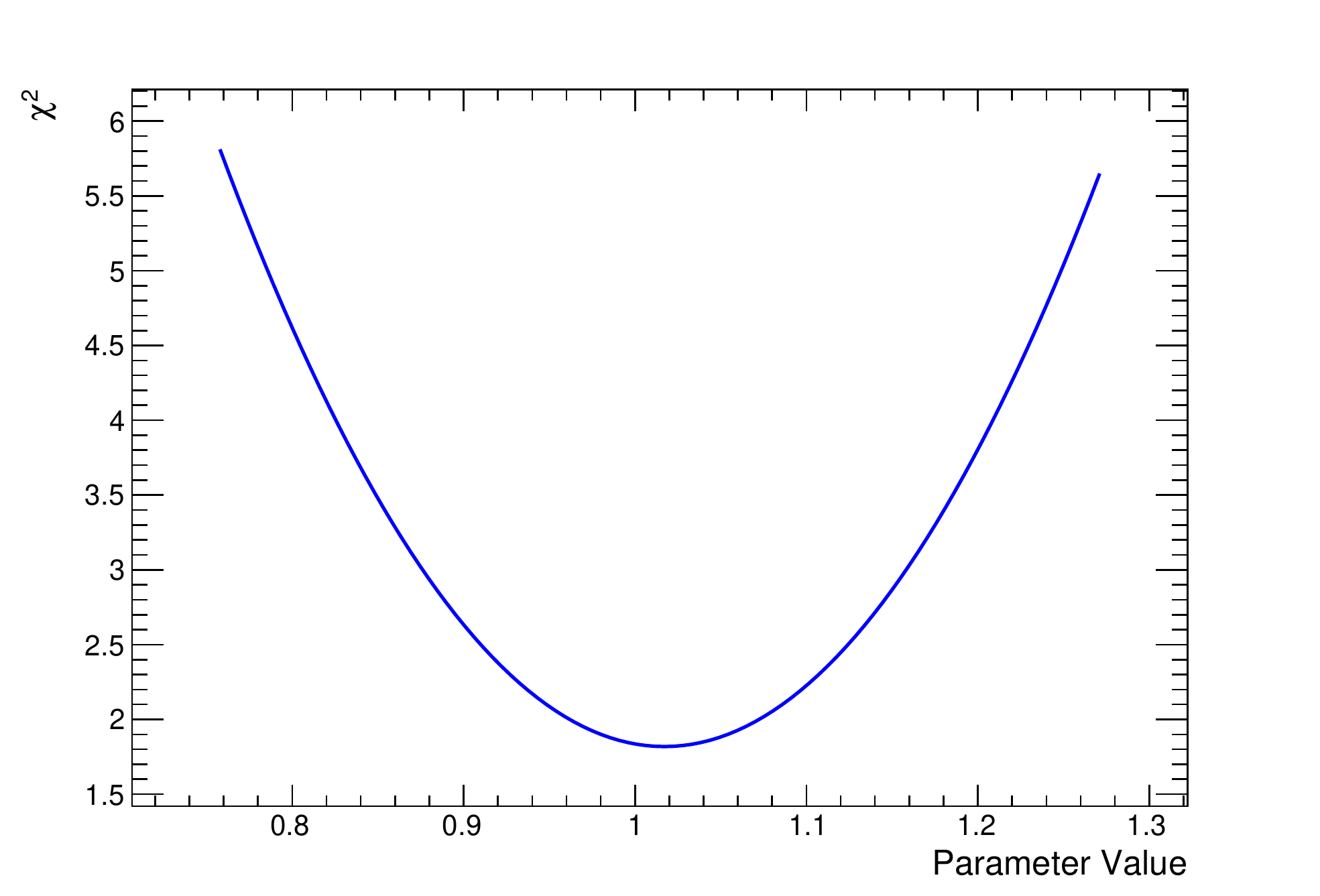}
        \caption{fReacHigh}
    \end{subfigure}
    
    \begin{subfigure}[t]{0.4\textwidth}
        \centering
        \includegraphics[width=\textwidth]{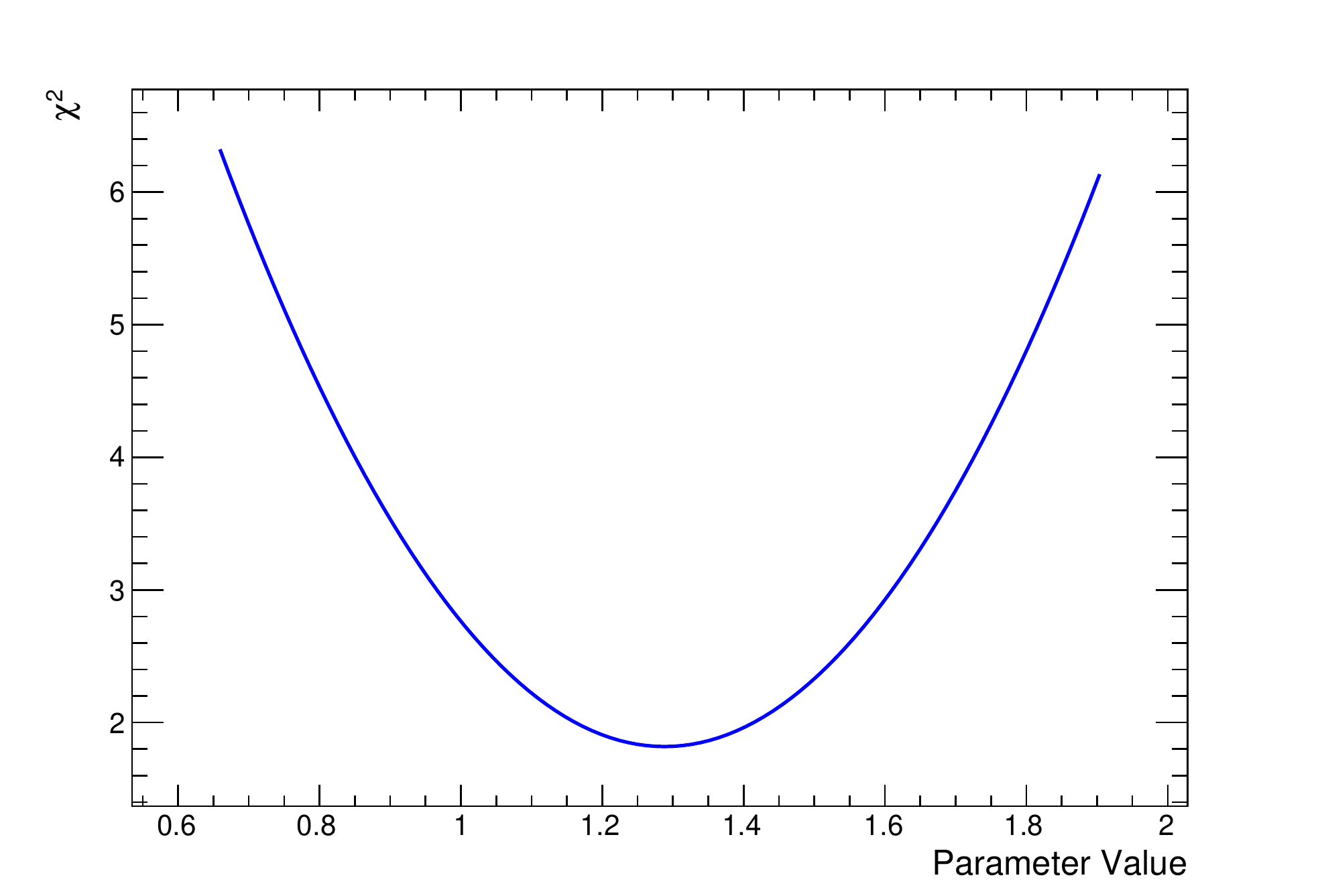}
        \caption{fAbs}
    \end{subfigure}
    ~
    \begin{subfigure}[t]{0.4\textwidth}
        \centering
        \includegraphics[width=\textwidth]{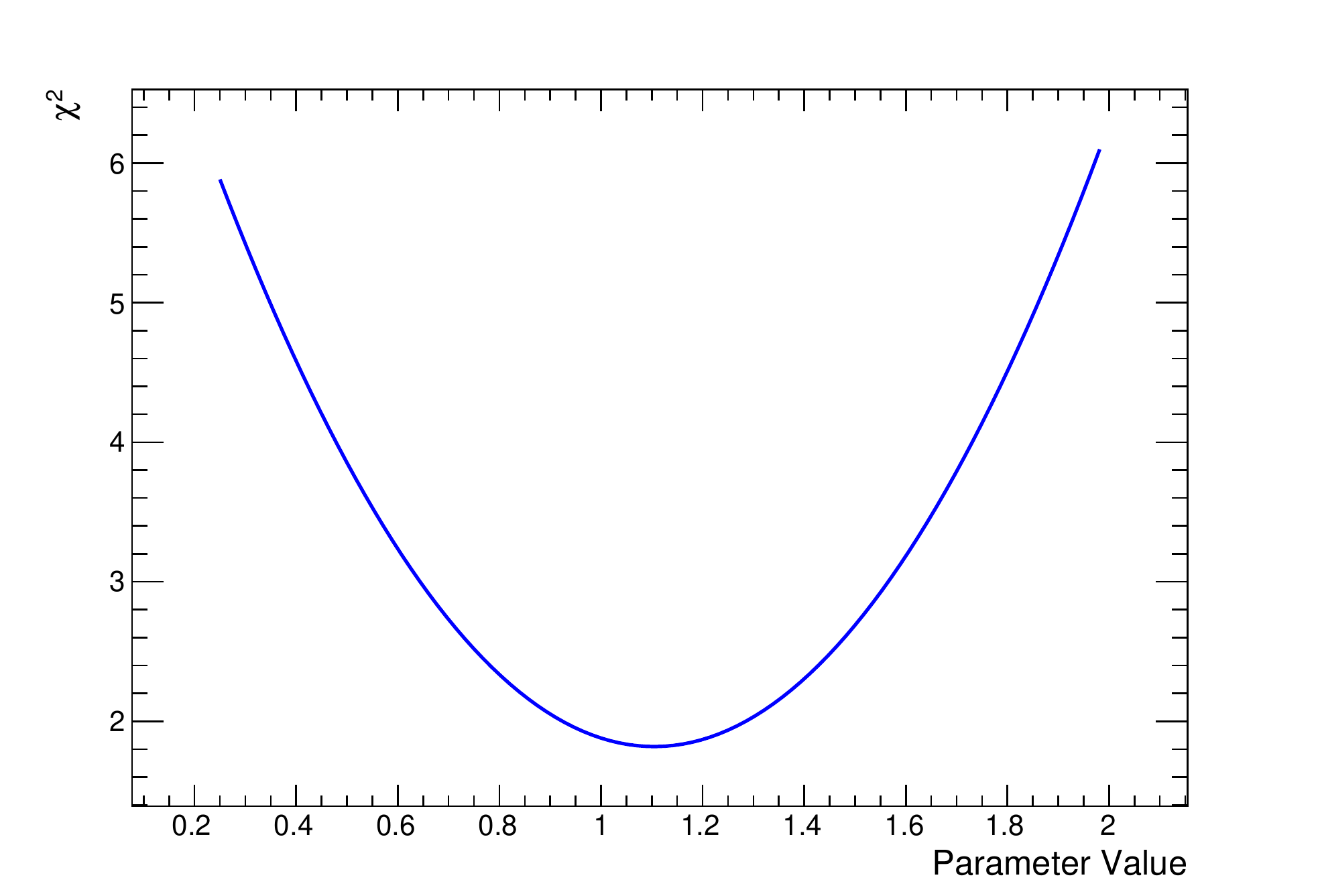}
        \caption{fCex}
    \end{subfigure}
    
    \begin{subfigure}[t]{0.4\textwidth}
        \centering
        \includegraphics[width=\textwidth]{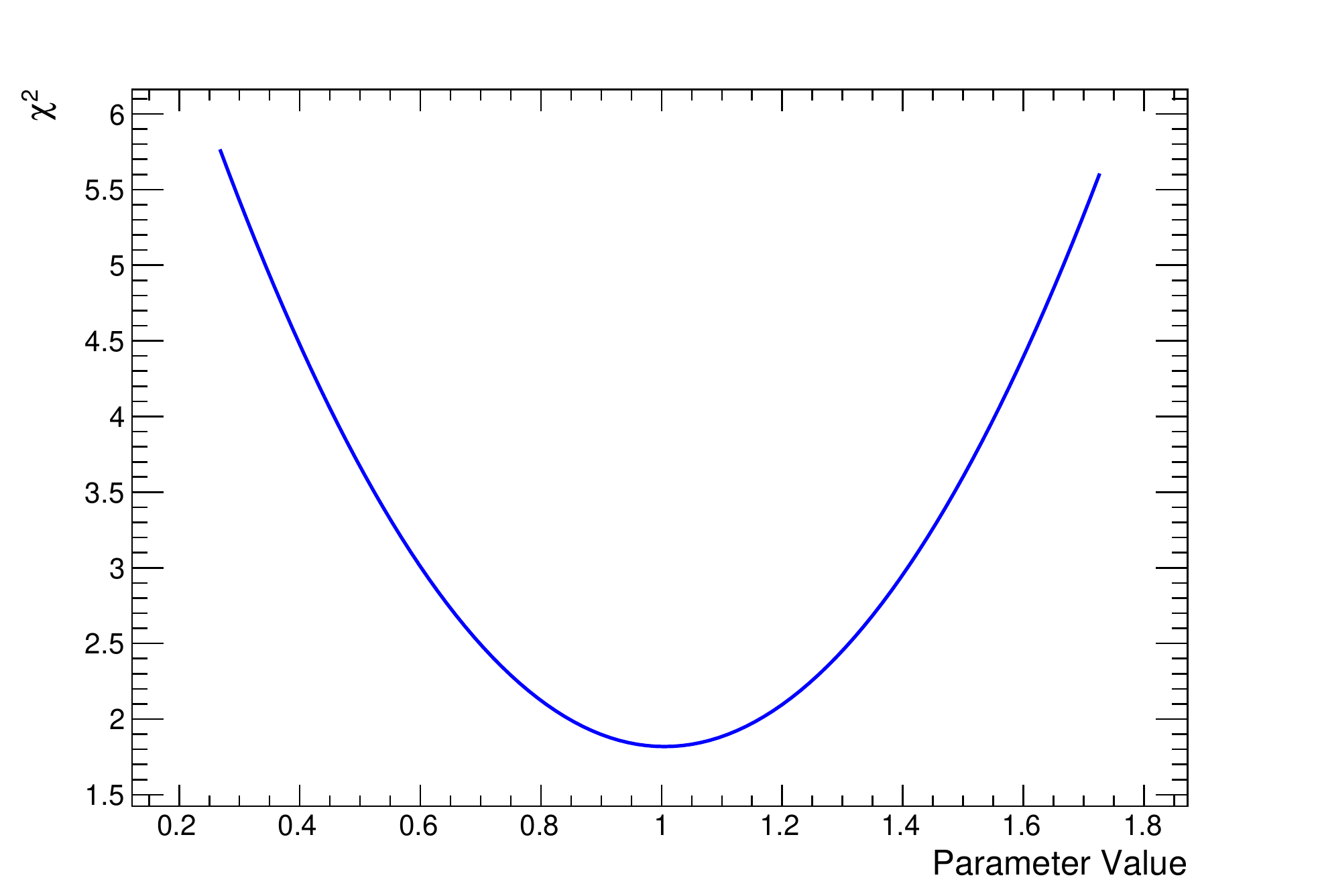}
        \caption{fInel}
    \end{subfigure}
    \caption{1-D parameter scans of the fit $\chi^2$ near the best-fit point.}
    \label{fig:scans}
   
\end{figure}

The results are shown in figures \ref{fig:pars_cov}--\ref{fig:scans}. Figure \ref{fig:pars_cov} shows the best fit values for the parameters and their covariance. Of note are the slightly larger covariances between the absorption and charge exchange parameters which are introduced due to including the combined absorption and charge exchange cross section data. In figure \ref{fig:fit}, cross section data used in the fit is compared to the nominal Geant4 prediction (blue line) and the best fit (black line) \& $\pm1\sigma$ error bands (red area). These figures highlight a current limitation in the framework as seen in the discontinuities in the best fit values of the cross sections and the error band sizes at the parameter transitions. Highly precise reaction data is included for momentum above 700 MeV/$c$. As such, this reduces the error on the high-momentum parameter as compared to the lower-momentum regions on the parameter affecting that region when compared to the middle region, which is fit to less accurate data. 
Finally, in figure \ref{fig:scans} are 1-D parameter scans of the fit $\chi^2$ around the best-fit point. All plots are generated by the fitting application and included in the output.

\clearpage


\section{Summary}
Geant4Reweight will serve the neutrino physics community by providing a framework that allows users to estimate and propagate systematic uncertainties related to hadron--nucleus interactions in Geant4. This paper describes the framework and its validation. Multiple experiments --- including MicroBooNE, NOvA, and ProtoDUNE-SP --- have begun to use this framework for uncertainty propagation in their analyses.



\acknowledgments
This material is based upon work supported by the U.S. Department of Energy (DOE), Office of Science, Office of Workforce Development for Teachers and Scientists, Office of Science Graduate Student Research (SCGSR) program. The SCGSR program is administered by the Oak Ridge Institute for Science and Education for the DOE under contract number DE‐SC0014664. Authors are also supported by DOE Award Number DE-SC0015903. We acknowledge the hard work of the Geant4 collaboration in the creation, documentation, and maintenance of the Geant4 software package, without which this work would not be possible. We also gratefully acknowledge a careful reading of this manuscript and useful comments by Wayne Repko.  




\begin{thebibliography}{99}


\bibitem{t2k_experiment}
T2K collaboration, \emph{The T2K Experiment}, \emph{Nucl. Instrum. Meth.} \textbf{A659} (2011) 106 [arXiv:1106.1238].

\bibitem{uboone}
MicroBooNE collaboration, \emph{Design and Construction of the MicroBooNE Detector}, \emph{JINST} \textbf{12} (2017) P02017 [arXiv:1612.05824].

\bibitem{nova}
NOvA collaboration, 
\emph{First measurement of muon-neutrino disappearance in NOvA}, \emph{Phys. Rev. D} \textbf{93}, (2016) 051104 [arXiv:1601.05037].

\bibitem{dune}
DUNE collaboration, 
\emph{Deep Underground Neutrino Experiment (DUNE), Far Detector Technical Design Report, Volume II: DUNE Physics}, {FERMILAB-PUB-20-025-ND} (2020) [arXiv:2002.03005].




\bibitem{geant4_1}
S. Agostinelli, \textit{et al.},
\emph{Geant4 -- a simulation toolkit} \emph{Nucl. Inst. Meth.} \textbf{A506} (2003) 250.

\bibitem{geant4_2}
Geant4 collaboration 
\emph{Geant4 developments and applications} \emph{IEEE Trans. Nucl. Sci.} \textbf{53} (2006) 270.

\bibitem{geant4_3}
Geant4 collaboration \emph{Recent developments in Geant4} \emph{Nucl. Inst. Meth.} \textbf{A835} (2016) 186.

\bibitem{T2K_Flux}
T2K collaboration, \emph{The T2K Neutrino Flux Prediction}, \emph{Phys. Rev. D} \textbf{87} (2013) 012001 [arXiv:1211.0469].

\bibitem{MINERvA_Flux}
MINERvA collaboration, \emph{Neutrino Flux Predictions for the NuMI Beam}, \emph{Phys. Rev. D} \textbf{87} (2016) 092005 [arXiv:1607.00704].

\bibitem{T2K_NuE_Appearance}
T2K collaboration
\emph{Evidence of Electron Neutrino Appearance in a Muon Neutrino Beam},
\emph{Phys. Rev. D} \textbf{88}, no.3, 032002 (2013)
doi:10.1103/PhysRevD.88.032002
[arXiv:1304.0841 [hep-ex]].



\bibitem{DobsonReweighting}
J.~Dobson and C.~Andreopoulos,
\emph{Propagating nu-interaction uncertainties via event reweighting},
Acta Phys. Polon. B \textbf{40}, 2613-2620 (2009).

\bibitem{DobsonThesis}
J.~Dobson
\emph{Neutrino Induced Charged Current $\pi^+$ Production at the T2K Near Detector}
Ph.D. Thesis, Imperial College London (2012)


\bibitem{g4rw_git}
Geant4Reweight GitHub Repository [\url{https://github.com/NuSoftHEP/Geant4Reweight}]

\bibitem{root}
R. Brun and F. Rademaker, \emph{ROOT — An object oriented data analysis framework}, 
\emph{Nucl. Instrum. Meth.} \textbf{A389} (2011) 81 - 86 [\url{https://doi.org/10.1016/S0168-9002(97)00048-X}].

\bibitem{geant4_app}
Geant4 collaboration \emph{Book For Application Developers} {Release 10.6} (2020) [\url{http://geant4-userdoc.web.cern.ch/geant4-userdoc/UsersGuides/ForToolkitDeveloper/BackupVersions/V10.6/fo/BookForToolkitDevelopers.pdf}].

\bibitem{geant4_phys}
Geant4 collaboration \emph{Physics Reference Manual} {Release 10.6} (2020) [\url{http://geant4-userdoc.web.cern.ch/geant4-userdoc/UsersGuides/PhysicsReferenceManual/BackupVersions/V10.6/fo/PhysicsReferenceManual.pdf}].

\bibitem{barashenkov}
V.S. ~Barashenkov and V.D. ~Toneev. \emph{High Energy interactions of particles and nuclei with nuclei (In
russian)}. 1972.

\bibitem{glauber}
R.J. ~Glauber. \emph{High Energy Physics and Nuclear Structure}. Plenum Press, NY, edited by s. devons edition,
1970.

\bibitem{gheisha}
H. ~Feseleldt, report PITHA-85/02, RWTH Aachen, 1985.


\bibitem{protodune_tdr}
DUNE collaboration, \emph{The Single-Phase ProtoDUNE Technical Design Report} (2017), [arXiv:1706.07081].

\bibitem{protodune_performance}
DUNE collaboration, \emph{First results on ProtoDUNE-SP liquid argon time projection chamber performance from a beam test at the CERN Neutrino Platform}
(2020), [arXiv:2007.06722].

\bibitem{larsoft}
E. D. Church. \emph{LArSoft:A Software Package for Liquid Argon Time Projection Drift Chambers}, (2013) [arXiv:1311.6774].

\bibitem{DUET}
E. S. Pinzon Guerra, \textit{et al.}, \emph{Measurement of \ensuremath{\sigma_{ABS}} and \ensuremath{\sigma_{CX}} of \ensuremath{\pi^{+}} on carbon by the Dual Use Experiment at TRIUMF (DUET)}, \emph{Phys. Rev. C} \textbf{95} (2017) 045203 [arXiv:1611.05612].

\bibitem{Allardyce}
B. W. Allardyce, \textit{et al.}, \emph{Pion reaction cross sections and nuclear sizes}, \emph{Nuclear Physics A} \textbf{209} (1973) 01 - 51.

\bibitem{Ashery2173}
D. Ashery, \textit{et al.}, \emph{True absorption and scattering of pions on nuclei}, \emph{Phys. Rev. C} \textbf{23} (1981) 2173 - 2185.

\bibitem{Ashery946}
D. Ashery, \textit{et al.}, \emph{Inclusive pion single-charge-exchange reactions}, \emph{Phys. Rev. C} \textbf{30} (1984) 946 - 951.

\bibitem{Bellotti}
E. Bellotti, \textit{et al.}, \emph{Positive-pion absorption by C nuclei at 130 MeV}, \emph{Nuov Cim A} \textbf{18} (1973) 75.

\bibitem{Levenson}
S. M. Levenson, \textit{et al.}, \emph{Inclusive pion scattering in the $\Delta$(1232) region}, {Phys. Rev. C} \textbf{28} (1983) 326.

\bibitem{Meirav}
O. Meirav, \textit{et al.}, \emph{Low energy pion-nucleus potentials from differential and integral data}, \emph{Phys. Rev. C} \textbf{40} (1989) 843.

\bibitem{Navon}
I. Navon, \textit{et al.}, \emph{True absorption and scattering of 50 MeV pions}, \emph{Phys. Rev. C} \textbf{28} (1983) 2548.

\bibitem{Saunders}
A. Saunders, \textit{et al.}, \emph{Reaction and total cross sections for low energy $\pi^+$ and $\pi^-$ on isospin zero nuclei}, \emph{Phys. Rev. C} \textbf{53} (1996) 1745.









\end{thebibliography}
\end{document}